\documentclass[12pt]{iopart}

\newcommand{\dd}{\mathrm{d}}

\usepackage[compat=1.1.0]{tikz-feynman}
\usepackage{hyperref}
\usepackage{amsmath}
\usepackage[caption=false]{subfig}
\usepackage{lineno}
\usepackage{comment}
\usepackage{cite}
\usepackage{units}
\usepackage{feynmp-auto}

\begin{document}

\title[]{A sterile-neutrino search using data from the MicroBooNE liquid-argon time projection chamber performed in an undergraduate teaching laboratory}


\author{John Waiton$^1$, Joseph Bateman$^{1}$, Justin J. Evans$^1$, Ole Gunnar Finnerud$^1$, Elena Gramellini$^1$, Roxanne Guenette$^1$, Pawel Guzowski$^1$, Aleksander Kedziora$^1$, Stefan S\"oldner-Rembold$^2$}
\address{$^1$University of Manchester, Manchester M13 9PL, UK}
\address{$^2$Imperial College of Science, Technology and Medicine, London SW7 2BZ, UK}

\ead{elena.gramellini@manchester.ac.uk}
\vspace{10pt}
\begin{indented}
\item[]July 2025
\end{indented}

\begin{abstract}

Fundamental particle physics is a key part of an undergraduate physics curriculum, but can be challenging to incorporate into teaching laboratories.
We present an undergraduate laboratory experiment that enables students to work with real data from the MicroBooNE liquid-argon time projection chamber to search for the existence of a sterile neutrino: a new, fourth, neutrino state. This search has galvanized physicists for decades, as the existence of sterile neutrinos could play a critical role in understanding fundamental processes in the early universe, provide viable dark matter candidates, and offer a natural explanation for the origin of neutrino masses. From an experimental point of view, anomalies observed in several neutrino experiments provide tantalizing hints of their existence. The analysis presented here, whilst based on real data, has been adapted as a pedagogical tool for an undergraduate teaching laboratory, in which students are asked to develop an understanding of neutrino oscillation theory, analyse a MicroBooNE data sample, and apply statistical methods to search for a potential sterile neutrino. The analysis incorporates machine learning techniques to improve event classification. Students are encouraged to explore these and other methods to optimize their results.

\end{abstract}
\tableofcontents

\section{Introduction}

Teaching laboratories are a core component of an undergraduate physics degree. These laboratories offer multiple educational benefits~\cite{ref:NatureLaboratoryPurpose,ref:GoalsOfIntroPhysicsLab,ref:TransformingLab,ref:LearningInTheLab}. They provide an opportunity to reinforce concepts taught in lecture courses through practical application, offer hands-on training in experimental techniques, and are often a student’s first exposure to statistical analysis~\cite{ref:StatsInLab}.

At more advanced levels, teaching laboratories support open-ended, enquiry-based learning~\cite{ref:OpenEndedLab}, facilitate integrative learning by connecting concepts from across the curriculum~\cite{ref:IntegrativeLearning}, and provide experience of working within an academic research environment~\cite{ref:ResearchLikeLabs}.

Incorporating particle physics into an undergraduate laboratory can be challenging, as much of the research in this field is conducted within large international collaborations at major facilities such as CERN~\cite{ref:CERN} and Fermilab~\cite{ref:Fermilab}. Nonetheless, since particle physics is a core component of most undergraduate physics programmes, it is important to integrate it into the accompanying teaching laboratory. In addition to reinforcing topics from the particle physics, relativity, and quantum mechanics syllabi, experimental particle physics introduces students to a wide range of computational and statistical techniques, and provides insight into cutting-edge research at the frontiers of physics.

In this paper, we present a particle physics data-analysis experiment, ``\textit{Study of muon-neutrino oscillations at the MicroBooNE experiment}", originally developed for the third-year undergraduate teaching laboratory at the University of Manchester~\cite{ole}. The experiment has been designed to be easily portable and has recently been implemented as a third-year laboratory project at Imperial College London.
These third-year laboratories are mandatory components of the physics curriculum for both Bachelor's and Master's degrees in Physics, and the experiment described here is one of several options available to students. This experiment offers students the opportunity to work with data from the MicroBooNE liquid-argon time projection chamber~\cite{ref:MicroBooNEDetector}, a major neutrino detector at Fermilab in Chicago, which was exposed to a beam of neutrinos produced by the Fermilab accelerator complex.
The dataset provided to students, although reduced in scale, closely resembles that used by researchers within the MicroBooNE collaboration. Students work on the project in pairs, for two days per week over a eight-week period.

Through this experiment, students are introduced to the quantum mechanical framework underlying neutrino oscillations and their phenomenological application to the propagation of neutrinos in an accelerator-produced beam. They then develop Python code to search for evidence of neutrino oscillations in the data provided. In parallel, they are given simulated data with which they develop computational techniques to isolate signal-enhanced samples. These signal-isolation methods are subsequently applied to the real data.
With access to research-level data and simulation, this aspect of the project is highly open-ended and allows students to incorporate machine-learning (ML) techniques if they wish.

Once students have isolated signal-enhanced samples, they then modify the simulated data to incorporate neutrino oscillation phenomena, enabling them to apply statistical techniques to compare the simulation with the real data and extract limits on the parameters of the oscillation model.
Data visualisation skills are also developed through the experiment, as students are required to produce graphs at each stage to evaluate their methods and to communicate their progress to laboratory demonstrators.

A significant advantage of this student project is its ease of implementation at low cost, requiring only a PC  and the capacity to develop and run Python code. During the 2024--25 academic year at Manchester, we ran this experiment with five pairs of students simultaneously, supported by a rotating team of three demonstrators. Practical details on running the experiment—including instructions for obtaining the code, data and simulation samples, as well as the laboratory script and reference solutions—are provided in Section~\ref{sec:Tech}.



This paper describes the structure of the laboratory experiment in detail, beginning with an introduction to the theoretical context and an overview of the liquid-argon technique for neutrino detection. We then present the methods used to isolate a sample of signal-like events from the MicroBooNE data and discuss the statistical hypothesis-testing techniques employed to extract limits on the neutrino oscillation parameters.

After completing the core part of the experiment, students are encouraged to explore potential improvements—such as those outlined in section~\ref{sec:extend}—with guidance from laboratory demonstrators, and to independently extend their analysis beyond the material described in this paper.

\subsection{Accessing the laboratory material}\label{sec:Tech}

This laboratory requires a reasonable level of python knowledge---the ability to install and use packages such as \texttt{pandas} and \texttt{numpy}, as well as the ability to complete troubleshooting---but can be attempted in any language.
Computationally, \unit[1]{GB} of storage space is \textbf{required} and a minimum of \unit[8]{GB} of memory is recommended for running this laboratory comfortably.

The laboratory is available on github in two forms: a public \href{https://github.com/jwaiton/uboone_lab_student_copy}{student copy} and a private \href{https://github.com/jwaiton/uboone_lab_teacher_copy}{teacher copy} containing solutions. For access to the teacher copy please contact \textbf{elena.gramellini@manchester.ac.uk}. The student copy includes the most up-to-date version of the lab script and code base. Data is publicly available at the following  \href{https://zenodo.org/records/17174393}{link}.  
\section{Theoretical context---neutrino oscillations}\label{sec:neu}

Neutrinos are fundamental particles and are the most common matter particle in the universe.
In particle physics, the Standard Model is the theoretical framework that describes the interactions of the fundamental particles.  Within this model, neutrinos (denoted as $\nu$) are classified as leptons---a type of elementary particle that does not participate in the strong nuclear force. Neutrinos are electrically neutral and possess a spin quantum number of $1/2$, indicating they have an intrinsic form of angular momentum characteristic of matter particles.

\subsection{A short history of neutrino oscillations}
The first experimental evidence for the existence of neutrinos emerged in the early 20th century from studies of $\beta$ decay. In these studies, a neutron ($n$) within a nucleus was observed to decay into a proton ($p$) and an electron ($e^{-}$). Contrary to expectations, the emitted electron exhibited a continuous energy spectrum rather than a discrete energy value, seemingly violating the principle of energy conservation~\cite{chadwick}.

In 1930, Wolfgang Pauli proposed a solution to this problem, postulating that in $\beta$ decay the nucleus simultaneously emits a particle with no electric charge and a spin quantum number of $1/2$, in order to conserve angular momentum~\cite{Pauliletter}. The $\beta$ decay process could therefore be described as 
$n \rightarrow p + e^{-} + \bar{\nu}_e$. 
Enrico Fermi incorporated Pauli’s proposed neutrino into his seminal 1934 theory of $\beta$ decay~\cite{ref:Fermi,ref:FermiEnglish}.

The existence of the neutrino was experimentally confirmed only 20 years later---in 1956---by Los Alamos physicists Frederick Reines and Clyde Cowan, who detected electron antineutrinos~\cite{reines}. Over the course of the twentieth century, three generations of neutrinos (\( \nu_e \), \( \nu_\mu \), \( \nu_\tau \)) were discovered, each associated with one of the charged lepton flavors ($e$, $\mu$, $\tau$). The most recent discovery was that of the tau neutrino ($\nu_\tau$), observed by the DONUT Collaboration in 2000~\cite{DONUT}.

Shortly after the experimental discovery of neutrinos, unexpected discrepancies in their behaviour gave rise to a puzzle known as the {\it solar neutrino problem}. In the late 1960s, the Homestake experiment first reported a significant deficit in the flux of solar electron neutrinos detected on Earth compared to predictions from the Standard Solar Model~\cite{HOMESTAKE_fullpaper,ref:StandardSolarModel}.
The Sun’s nuclear fusion processes produce electron neutrinos at a rate determined by its luminosity, and due to their extremely weak interactions with matter, these neutrinos were expected to escape the Sun and reach Earth largely unimpeded. However, multiple experiments consistently measured less than half of the predicted flux~\cite{ref:SAGE,ref:GALLEX,ref:Kamiokande,ref:SuperKamiokandeSolar}.

The solution to this discrepancy is the phenomenon of {\bf neutrino oscillations}, first proposed by Bruno Pontecorvo in 1967~\cite{Pontecorvo} and later refined by Maki, Nakagawa, and Sakata~\cite{ref:MNS}. In this process, a neutrino of one flavour  (\( \nu_e \), \( \nu_\mu \), or \( \nu_\tau \)) 
 can transform into another flavour as it travels. Consequently, some of the electron neutrinos produced by the Sun convert into muon or tau neutrinos on their way to Earth, resulting in an apparent deficit in detectors that are sensitive only to electron-neutrino interactions.
Neutrino oscillations have since been confirmed by several experiments, including Super-Kamiokande~\cite{Fukuda_1998} and SNO~\cite{ref:SNO}, culminating in the award of the 2015 Nobel Prize in Physics to Takaaki Kajita and Arthur McDonald for ``the discovery of neutrino oscillations, which shows that neutrinos have mass"~\cite{nobel}.

\subsection{Neutrino mixing}\label{sec:oscillation}
Neutrino oscillations are a quantum mechanical phenomenon that determines the probability of detecting a neutrino of a given flavour depending on the neutrino's energy and distance travelled. The three \textbf{neutrino flavour eigenstates}, $\nu_e$, $\nu_\mu$, and $\nu_\tau$, can, in a formulation called {\bf neutrino mixing}, be written as linear superpositions of \textbf{mass eigenstates}, $\nu_{1}$, $\nu_{2}$, and $\nu_{3}$, with definite masses $m_1$, $m_2$, and $m_3$. The $3\times 3$ Pontecorvo–Maki–Nakagawa–Sakata (PMNS) matrix, $U^{\mathrm{PMNS}}$,
parametrises the transformation (or rotation) between the flavour and mass eigenstates~\cite{pdg}:
\begin{equation} \label{eqn1}
\begin{pmatrix}
  \nu_{e} \\
  \nu_{\mu} \\
  \nu_{\tau} \\
\end{pmatrix}
=
U^{\mathrm{PMNS}}\begin{pmatrix}
  \nu_{1}\\\nu_{2}\\\nu_{3}
\end{pmatrix}
= 
\begin{pmatrix}
U_{e1} \; \; U_{e2}\; \; U_{e3}\\
U_{\mu1} \; \; U_{\mu2}\; \; U_{\mu3}\\
U_{\tau1} \; \; U_{\tau2}\; \; U_{\tau3}\\
\end{pmatrix}\begin{pmatrix}
  \nu_{1}\\\nu_{2}\\\nu_{3}
\end{pmatrix}.
\end{equation}
This complex matrix is parameterised with three {\bf mixing angles}, \( \theta_{12} \), \( \theta_{13} \), and \( \theta_{23} \), and a complex phase. 
The relationship between the mass eigenstates is given by two independent {\bf mass splittings}, $\Delta m^2_{21}=m_2^2-m_1^2$ and $\Delta m^2_{32}=m_3^2-m_2^2$. 



\subsection{Neutrino oscillations in a two-flavour model}\label{sec:TwoFlavourModel}

\begin{figure}
    \centering
    \includegraphics[width=0.8\linewidth]{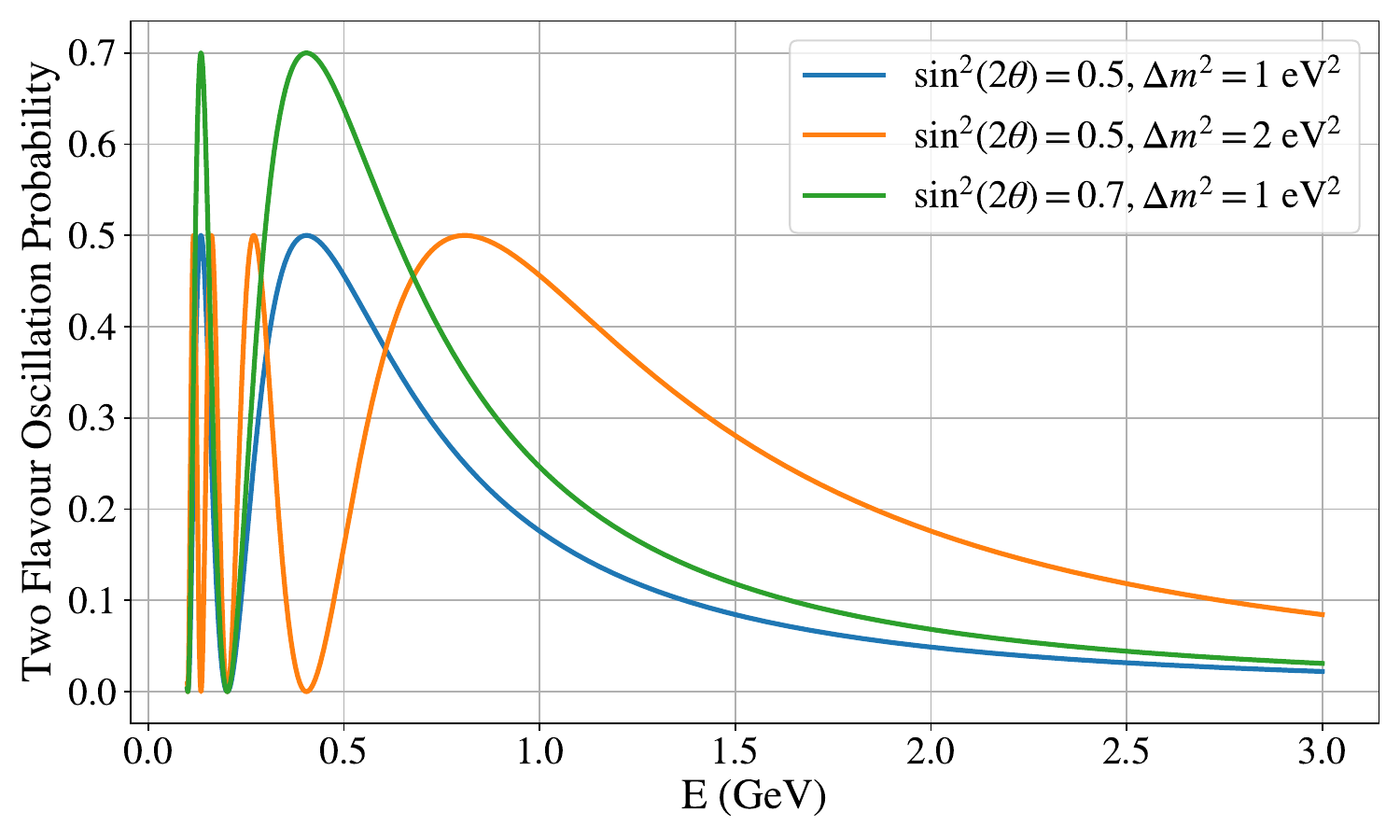}
    \caption{The two-flavour probability for neutrino flavour change, $P(\nu_\alpha\rightarrow\nu_\beta)$, from equation~(\ref{eq:natty}), shown as a function of neutrino energy, for a distance $L=\unit[468.5]{m}$, which corresponds to the location of the MicroBooNE detector in the Fermilab neutrino beam. The probability is shown for different values of the mixing angle $\theta$ and the mass splitting $\Delta m^2$.}
    \label{fig:OscProb}
\end{figure}

The three-flavour model can be simplified to a two-flavour model with two flavour eigenstates \mbox{($\nu_{\alpha}$, $\nu_{\beta}$)} and two mass eigenstates \mbox{($\nu_{1}$, $\nu_{2}$)}~\cite{Diaz_2020}. The PMNS matrix  is therefore reduced to a real $2 \times 2$ unitary rotation matrix, with a single mixing angle $\theta$:
\begin{equation} \label{eqn2}
\begin{pmatrix}
  \nu_{\alpha} \\
  \nu_{\beta}\\
\end{pmatrix}
=
\begin{pmatrix}
U_{\alpha1} \;\; U_{\alpha2} \\
U_{\beta1} \;\; U_{\beta2}
\end{pmatrix}
\begin{pmatrix}
  \nu_{1}\\\nu_{2}
\end{pmatrix}
=
\begin{pmatrix}
  \;\;\;\cos\theta \;\; \sin \theta\\ -\sin\theta \;\; \cos\theta
\end{pmatrix}\begin{pmatrix}
  \nu_{1}\\\nu_{2}
\end{pmatrix}
\end{equation}
Using the two-flavour model, students can more easily derive the probability $P(\nu_\alpha \rightarrow \nu_\beta)$ of oscillation from a neutrino of flavour $\nu_\alpha$ to one of flavour $\nu_\beta$. We consider a neutrino, $|\nu(t,{\bf x})\rangle$, produced at time $t=0$ and position ${\bf x}=0$ in a $\nu_\alpha$ eigenstate,
\begin{equation}
    |\nu(0,0)\rangle=|\nu_\alpha\rangle = \cos\theta\,|\nu_{1}(0,0)\rangle +\sin\theta\,|\nu_{2}(0,0)\rangle.\label{eq:NeutrinoAtTimeZero}
\end{equation}
The time evolution of the mass eigenstates is described via the plane-wave solution for the time-dependent Schr\"odinger equation,
\begin{equation}
|\nu_j(t, \textbf{x})\rangle = e^{-i(E_j t-\mathbf{p_j}\cdot \mathbf{x})}|\nu_j(0,0)\rangle,\label{eq:TimeVaryingMassStates}
\end{equation}
with  $E_{j}$ being the energy and $\mathbf{p_{j}}$ the momentum of the neutrino mass state $j$.
We use natural units, where $\hbar=c=1$, and treat the neutrinos as ultra-relativistic particles such that $|{\bf p}_1|\approx|{\bf p}_2|\equiv E$ and $t\approx x \equiv L$, where $L$ is the distance the neutrino has travelled from the point of production, such that 
\begin{equation}
    E_j t-\mathbf{p_j}\cdot \mathbf{x}\approx\frac{m_jL}{2E}.
\end{equation}
Using appropriate units, the probability of oscillation between two states can then be written as
\begin{equation} \label{eq:natty}
    P(\nu_\alpha \rightarrow \nu_\beta) = |\langle\nu_\beta|\nu(t,\mathbf{x})\rangle|^2 = \textrm{sin}^2(2\theta)\,\textrm{sin}^2\left(1.27\frac{\Delta m^2 (\textrm{eV}^2) L\textrm{(km)}}{E\textrm{(GeV)}}\right),
\end{equation}
where $\Delta m^2 = m_2^2 - m_1^2$ is the mass splitting between the two mass eigenstates. For the study of muon-neutrino disappearance later in this project, the corresponding two-flavour {\bf survival probability} 
\begin{equation}\label{eq:TwoFlavourSurvivalProb}
    P(\nu_\alpha \rightarrow \nu_\alpha) = 1-P(\nu_\alpha\rightarrow\nu_\beta) 
\end{equation}
is used. This flavour-change probability is shown in figure~\ref{fig:OscProb} for a baseline of $L=\unit[468.5]{m}$, corresponding to the location of the MicroBooNE detector in the Fermilab neutrino beam.
As illustrated in figure~\ref{fig:OscProb}, the neutrino flavour-change probability is oscillatory, with a frequency governed by the mass splitting $\Delta m^2$ and an amplitude governed by the mixing angle $\theta$. Neutrino oscillations
therefore represent direct evidence for non-zero neutrino masses. 

Students are expected to be able to derive equation~(\ref{eq:natty}), starting from equations~(\ref{eq:NeutrinoAtTimeZero}) and~(\ref{eq:TimeVaryingMassStates}).
They are asked to plot the oscillation probability given in equation~(\ref{eq:natty}), and to vary the parameters to understand how this probability depends on $\theta$, $\Delta m^2$, $L$, and $E$.

\subsection{Neutrino oscillations with more than two flavours}
\label{sec:3+1neut}

We previously discussed that three-neutrino phenomenology is characterised by two distinct mass splittings. The corresponding mixing angles and mass splittings have been measured extensively across many experiments~\cite{daya13, superkangles, nova_angles, t2k_angles, icecube_angles, ref:SNOMostRecent, ref:KamLANDMassSplitting}. Combined fits of all measurements improve both precision and accuracy, and provide a means of testing consistency~\cite{nufit, globalfit}.

The two mass splittings have measured values of
$\Delta m^2_{21}=\unit[(7.5\pm 0.2)\times 10^{-5}]{eV^2}$ and $\Delta m^2_{31}=\unit[\pm (2.5\pm 0.2)\times 10^{-3}]{eV^2}$~\cite{nufit}. 
The two mass splittings give rise to two characteristic oscillation wavelengths of $L/E\approx\mathcal{O}(\unit[10^{5}]{km/GeV})$ and $L/E\approx\mathcal{O}(\unit[10^{3}]{km/GeV})$ (see equation (\ref{eq:natty})).
Because the two mass splittings---and therefore the frequencies and wavelengths of oscillations between flavour eigenstates---differ significantly, the three-flavour oscillation framework can be approximated as a superposition of independent two-flavour oscillations. This validates the simplified model discussed previously.

\subsection{A fourth neutrino?}
The general consensus that a three-flavour model provides an optimal description of observed neutrino phenomena has been called into question by results from experiments such as MiniBooNE~\cite{MiniBooNE2021} and LSND~\cite{ref:LSND}. These experiments report anomalous 
 $\nu_\mu\rightarrow\nu_e$ transitions over $L/E\approx\mathcal{O}(\unit[1]{km/GeV})$, referred to as {\bf short-baseline} oscillations
 as they occur over much shorter distances than those associated with standard neutrino oscillations.
Such oscillations at these $L/E$ values cannot be accommodated within the standard three-flavour framework and would require the existence of at least one additional mass splitting of $\mathcal{O}(\unit[1]{eV^2})$.

The simplest model that incorporates an additional mass splitting is the ``$3+1$ neutrino model'', which postulates the existence of a fourth neutrino mass eigenstate $\nu_4$ and a corresponding fourth neutrino flavour eigenstate $\nu_s$~\cite{3plus1fit}. The PMNS matrix from equation~(\ref{eqn1}) is extended to become a 4×4 complex unitary matrix,
\begin{gather}
    \begin{pmatrix}
\nu_e \\ \nu_\mu \\ \nu_\tau \\ \nu_s
\end{pmatrix}
=
\begin{pmatrix}
U_{e1} & U_{e2} & U_{e3} & U_{e4} \\ U_{\mu1} & U_{\mu2} & U_{\mu3} & U_{\mu4} \\ U_{\tau1} & U_{\tau2} & U_{\tau3} & U_{\tau4} \\ U_{s1} & U_{s2} & U_{s3} & U_{s4}
\end{pmatrix}
\begin{pmatrix}
\nu_{1} \\ \nu_{2} \\ \nu_{3} \\ \nu_{4}
\end{pmatrix}.
\label{eqn: PMNS_4_flavour}
\end{gather}
This matrix is parameterised by six real mixing angles, $\theta_{ij}$, where $1\leq i<j\leq 4$, as well as three complex phases.

The fourth mass eigenstate is introduced with a mass $m_{4}\approx 1$~eV corresponding to $m_4^2 \approx\Delta m^2_{41}\approx\mathcal{O}(\unit[1]{eV^2})$, which is several orders of magnitude larger than $\Delta m^2_{21}$ and $\Delta m^2_{32}$, leading to oscillations on very short baselines. This allows the oscillation frequency to be governed by a single mass splitting, $\Delta m^2_{41}$, which is described by the two-flavour model.

The question remains as to why this fourth neutrino state has not been observed by other experiments that produce neutrinos. Measurements of $Z$-boson decays, for example, show that only three neutrino flavours couple to the $Z$ boson~\cite{ref:ZWidth}. The fourth neutrino flavour eigenstate is therefore assumed to be {\bf sterile}, meaning that it does not interact with Standard Model matter via the weak interaction. Its existence can only be inferred indirectly by studying oscillations of the other (detectable) neutrino flavours.

In later parts of this experiment, students interpret their muon-neutrino disappearance results in the context of the electron-neutrino appearance probability, $P(\nu_\mu\rightarrow\nu_e)$, to test the consistency of their findings with the observations reported by the LSND and MiniBooNE experiments. More details are given in section~\ref{sec:ThreeFlavourInterpretation}.

\section{Detecting neutrinos}\label{sec:detsim}

The data sample considered in this paper is recorded with the MicroBooNE neutrino detector~\cite{ref:MicroBooNEDetector}, which is an 85-tonne liquid-argon time projection chamber (LArTPC). 
This technology is widely used to detect neutrino interactions by current and future neutrino experiments such as LArIAT~\cite{LArIAT}, ICARUS~\cite{ref:ICARUS}, SBND~\cite{SBND}, and DUNE~\cite{ref:DUNETDR,ref:DUNEHDTDR,ref:DUNEVDTDR}.

\subsection{Generating a neutrino beam}

The MicroBooNE detector is exposed to a muon-neutrino beam, called the Booster Neutrino Beam (BNB), which is generated by directing a beam of protons with kinetic energy of \unit[8]{GeV} onto a beryllium target~\cite{bnbbeam}.
When the protons strike the target, they produce a cascade of hadrons, primarily charged pions ($\pi^\pm$) and kaons ($K^\pm$), which are focused into a beam using magnetic horns.
The neutrino production is dominated by the weak decay of charged pions, $\pi^+ \rightarrow \mu^+ \nu_\mu$ (and the charge conjugate process).
The decay products pass through a material region designed to absorb the majority of particles, while the weakly interacting neutrinos remain unaffected and can thus reach the detector.

\subsection{Neutrino interactions in argon}

Neutrinos cannot be detected directly as they carry no electric charge and do not emit any radiation.
Instead, their presence in the LArTPC detector is inferred by observing the products of their interactions with argon. When neutrinos undergo weak interactions with argon nuclei, they produce charged particles, which can then be detected.
These charged particles are key to understanding the neutrino's properties.

\begin{figure}[t]
  \centering
\includegraphics[angle=0,width=0.5\hsize]{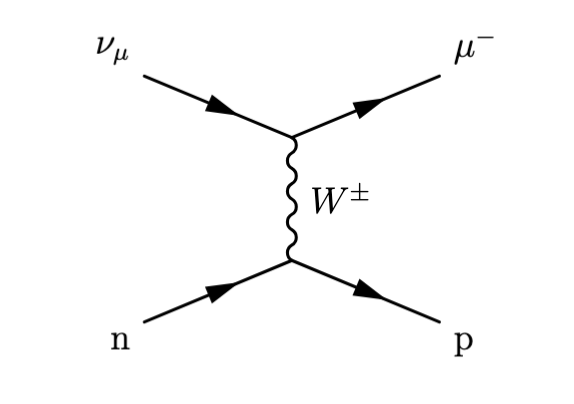}


\caption{Quasi-elastic scattering of a muon-neutrino ($\nu_{\mu}$) off a neutron ($n$). A charged current interaction, mediated by a $W^\pm$ boson, results in a muon ($\mu^-$) and proton ($p$).}
\label{fig1}
\end{figure}

An example neutrino interaction, shown in figure~\ref{fig1}, is a charged-current (CC) interaction.
In such events, a neutrino scatters off a neutron in the argon nucleus via exchange of a charged \( W \) boson, producing the corresponding charged lepton and a proton, with the flavour of the charged lepton identifying the flavour of the incoming neutrino.
Charged-current interactions constitute the signal of interest for this analysis.

The interactions can also result in more complex final states~\cite{FSI, multi-FSTs, neutral-current-resonance} comprising multiple pions and protons (see figure~\ref{fig: Event displays} for examples). Any possible CC signature is included in the signal definition for our analysis regardless of its final state.

Neutral current (NC) interactions are mediated by the $Z$~boson, which does not transport electric charge. They are considered to be background for this analysis as no charged lepton is produced and therefore the flavour of the incoming neutrino cannot be identified.

\subsection{A liquid-argon time projection chamber}
\begin{figure}[t]
\includegraphics[angle=0,width=0.9\hsize]{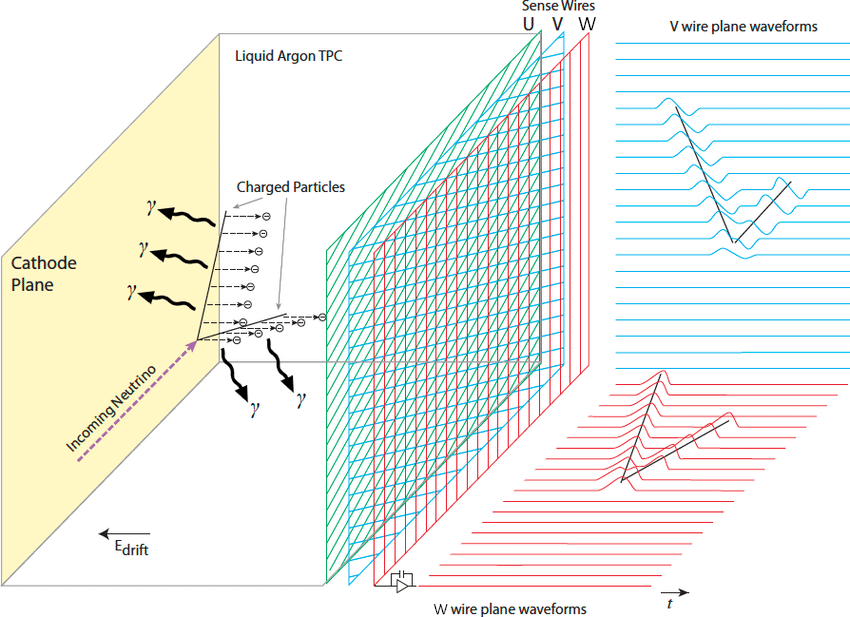}
\caption{A schematic describing the operation of a liquid-argon time projection chamber taken from Ref.~\cite{ref:MicroBooNEDetector}. An incoming neutrino interacts with an argon nucleon, releasing charged particles that produce scintillation light (denoted by $\gamma$) and ionisation electrons as they travel through the detector. The ionisation electrons drift onto the three wire planes (labelled U, V, and W), which interpret the change in current as signal (shown in the waveforms on the right).
}
\label{fig:tpc}
\end{figure}

A large cryostat houses a liquid-argon time projection chamber (LArTPC), as shown in figure~\ref{fig:tpc}. It functions as a large parallel-plate capacitor with a cathode and an anode. This chamber, immersed in liquid argon at cryogenic temperatures of \unit[87]{K}, forms the core of the neutrino interaction detection. Charged particles produced by neutrino interactions travel through the argon medium, generating both scintillation light and ionization electrons. The MicroBooNE LArTPC is \unit[2.3]{m} high, has a \unit[2.6]{m} width separating the anode and cathode, and is \unit[10.4]{m} long in the direction of the neutrino beam.

\begin{figure*}[t!]
    \centering
    \subfloat[]{
        \includegraphics[width=.45\linewidth]{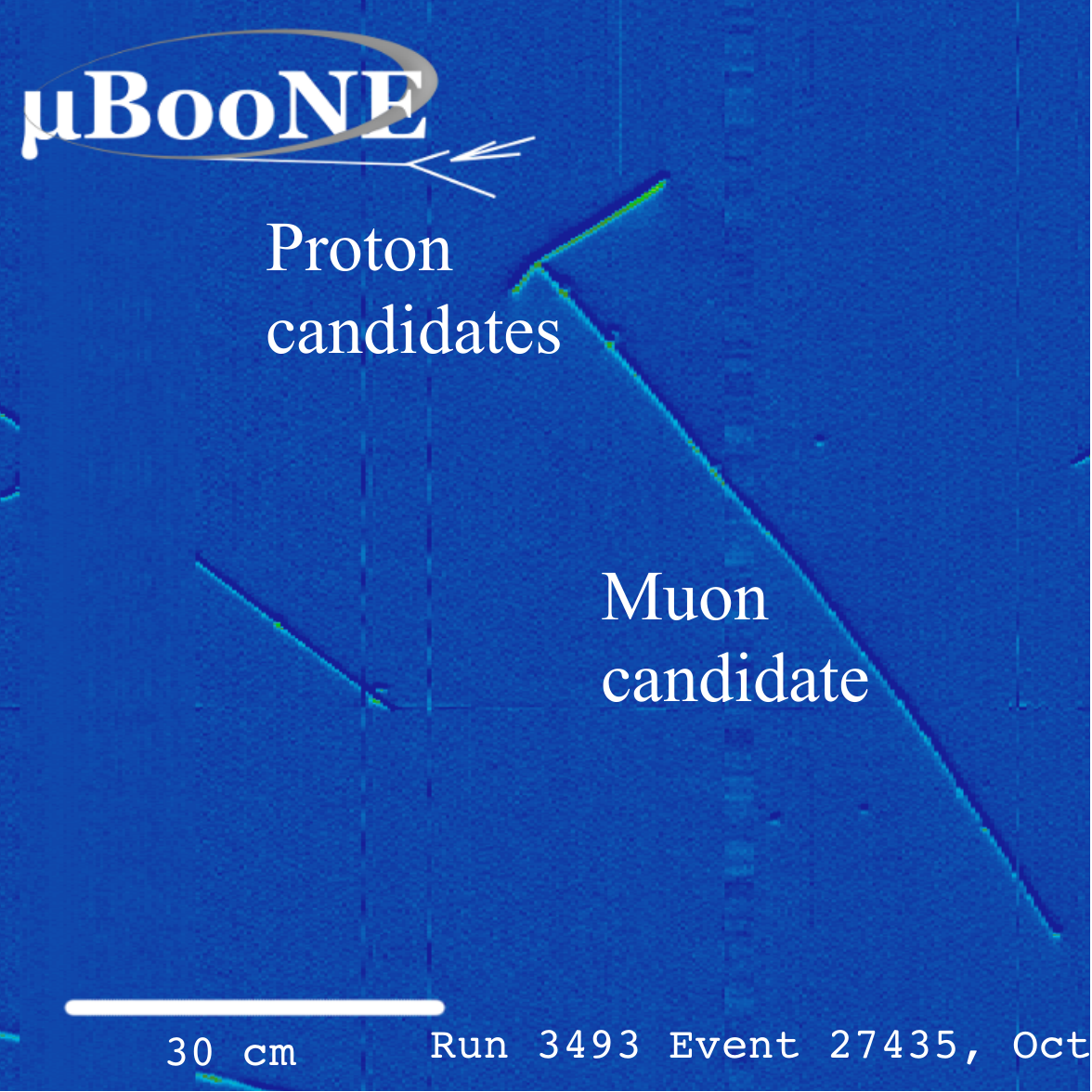}
        \label{subfig:a}}
    \hfill
    \subfloat[]{
        \includegraphics[width=.45\linewidth]{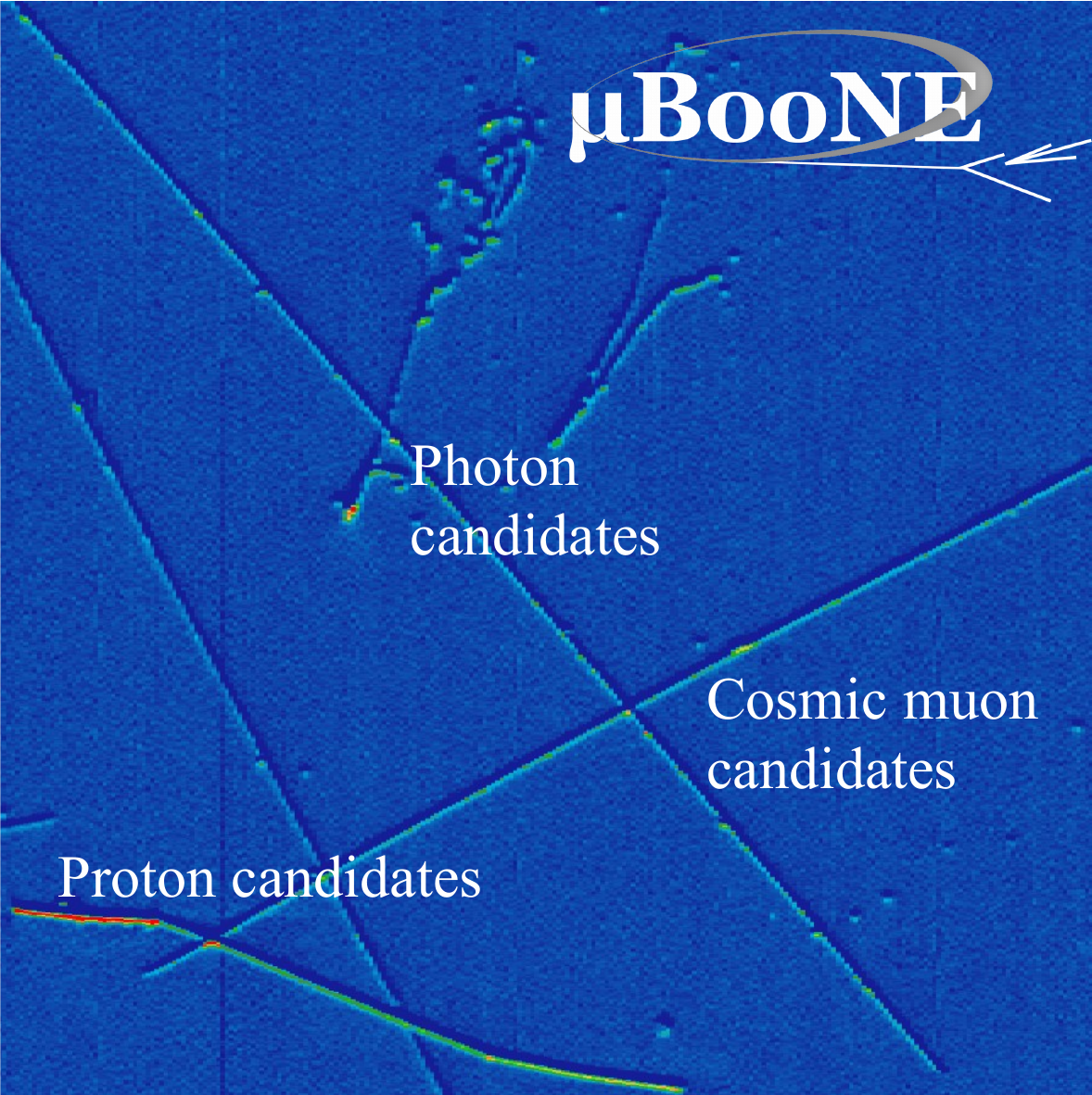}
        \label{subfig:b}
        }
    \hfill
    \centering
    \hfill
    \subfloat[]{

        \includegraphics[width=0.985\linewidth]{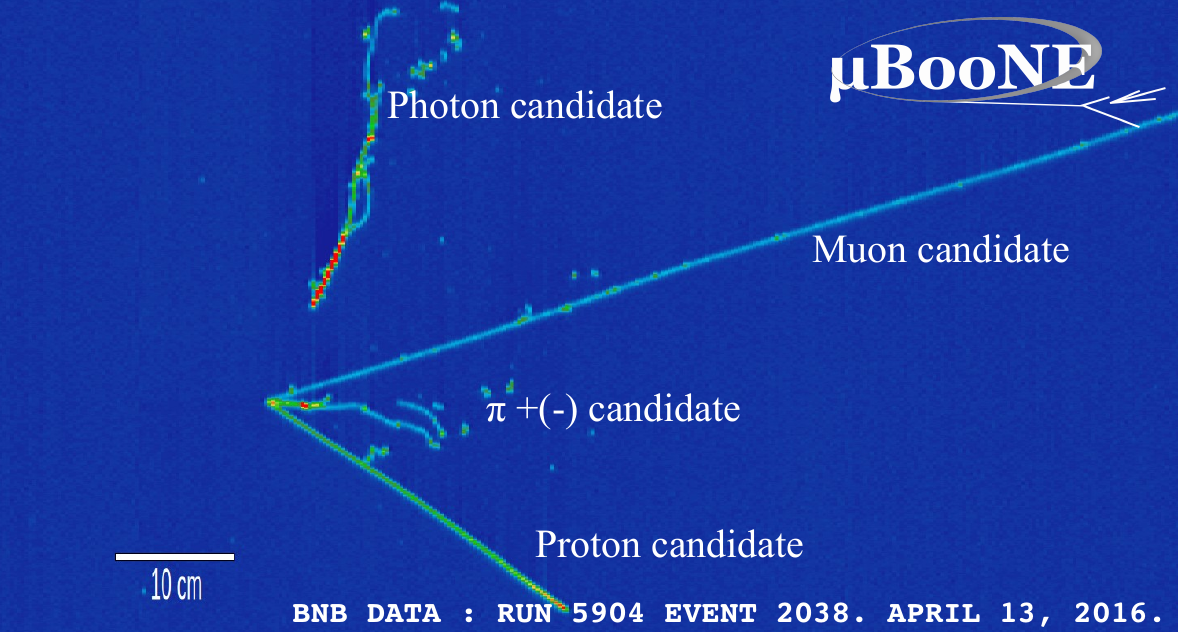}
        \label{subfig:c}}

    \caption{Event displays captured by MicroBooNE, labelled with candidate particles for the tracks and showers. Figures~\ref{subfig:a} and~\ref{subfig:c} show candidate charged current muon-neutrino interactions. Figure~\ref{subfig:b} is a neutral current interaction candidate. The colour map, from blue to red, indicates the amount of charge  (energy) deposited by the particles.
The $x$ axis shows the wire number and the $y$ axis shows the time, translated to a length scale.
    } 
    
    \label{fig: Event displays}
\end{figure*}

An electric field is applied across the argon to collect the ionization electrons, drifting them towards the anode comprising three wire planes oriented at different angles, as shown in figure~\ref{fig:tpc}. In MicroBooNE this field corresponds to $\unit[0.273]{kV\,cm^{-1}}$. 

The scintillation light is detected almost instantaneously by photon detectors such as photomultiplier tubes, typically placed behind the anode plane, while the charge signal arrives at the anode at a much longer time scale of milliseconds. By combining the location of the drifted charge, as reconstructed from the signals on the three wire planes, with timing information from the prompt scintillation light, the neutrino interaction can be reconstructed in three dimensions.
Examples of neutrino interactions reconstructed in the MicroBooNE detector are shown in figure~\ref{fig: Event displays}.

Since neutrinos can also interact with the detector materials surrounding the argon volume, resulting in incomplete tracking information, we define an inner region---known as the \emph{fiducial volume}---within the detector.
This ensures that only neutrino interactions with final-state particles that are contained in the detector are considered.
In the case of MicroBooNE, this volume is defined as \unit[5]{cm} from the sides of the detector, and only interactions within this volume are considered.

\subsection{Particle signatures in a liquid-argon time projection chamber}\label{subsec:detector physics}

As charged particles travel through the detector medium, they deposit energy at a rate that depends on their mass and momentum. This energy loss is characterised by the Bethe–Bloch equation, which describes the mean energy lost per unit distance, $\dd E/\dd x$, as particles traverse matter~\cite{pdg}. Measuring the specific energy loss, \( \dd E/\dd x \), allows for the identification of different particle species, as each exhibits a characteristic energy deposition profile. Particles are also identified topologically, with two distinct categories, \textbf{tracks} and \textbf{showers}.

\textbf{Tracks} are produced by charged particles such as protons, charged pions, and muons, which ionise atoms along approximately straight paths as they traverse the liquid argon. In contrast, \textbf{showers} are cascades generated by electrons, positrons, and photons. These electromagnetic showers arise because pair production---the creation of an electron–positron pair---is the dominant interaction mechanism for photons at the energies relevant to accelerator-based neutrinos.

The resulting electron–positron pairs are typically highly energetic and emit bremsstrahlung radiation as they propagate through the detector. This radiation can, in turn, produce further electron–positron pairs, initiating a cascade of secondary particles. As these charged particles travel through the argon, they ionise the medium, producing the characteristic signature of an electromagnetic shower.

Electromagnetic showers generated by photons or electrons can be distinguished based on their topology and $\dd E/\dd x$ profiles at the beginning of the shower. Photons, being electrically neutral, do not directly ionise the argon. Instead, they initiate electromagnetic showers through the aforementioned electron–positron pair production mechanism, which typically occurs at some distance from the reconstructed neutrino interaction point (the \textbf{vertex}). This results in a characteristic initial $\dd E/\dd x$ approximately twice that of a single electron or positron.

The excellent spatial and calorimetric resolution of MicroBooNE allows for the identification of both the displaced shower origin and the elevated initial energy deposition, enabling effective discrimination between photons and electrons. Examples of both showers and tracks are shown in figure~\ref{fig: Event displays}.

\subsection{Classification of neutrino interactions}\label{subsection: event classification}

Event displays are visual representations of detector data, used to illustrate the event topologies described in section~\ref{subsec:detector physics}. These displays are generated from the electronic signals read out by the detector (see figure~\ref{fig: Event displays}). The colour map indicates the amount of charge deposited by particles along their tracks or showers.

The $x$ axis represents the wire number, while the $y$ axis corresponds to the drift time. Since the wire spacing is known and the electron drift velocity towards the anode is approximately constant, both axes can be converted into spatial coordinates, allowing for a three-dimensional reconstruction of the particle trajectories within the detector.

Figure~\ref{subfig:a} shows a candidate event for a CC interaction of a muon neutrino, the signal process for this analysis. The neutrino interaction results in a muon and protons originating from the interaction vertex, following long, straight tracks through the argon. 
The muon track is typically significantly longer than that of a proton, and, as a minimum ionising particle, produces less ionisation per unit length (lower $\dd E/\dd x$).
Figure~\ref{subfig:b} shows a candidate NC interaction of a neutrino. In this event, proton candidates are produced at the interaction vertex with no outgoing charged lepton.

The display also shows two electromagnetic showers assumed to be the product of photons. Their trajectories suggest that they are caused by photons produced at the neutrino interaction vertex, as they can be traced back to a region close to the point labelled ``proton candidates.''
This is a typical signature of a neutral pion ($\pi^0$) immediately decaying into two photons.
Figure~\ref{subfig:b} also contains several high-energy muon candidates produced in the upper atmosphere by cosmic-ray interactions.
These cosmic muons typically traverse the entire detector volume.
Figure \ref{subfig:c} shows another example of a charged current muon-neutrino scattering event, with an additional photon shower offset from the interaction vertex.

Students are provided with a library of event displays.
They are asked to identify the various particle signatures, using their knowledge of the passage of charged particles through matter to justify their choices.
From the pattern of particle signatures in each event display, they are asked to identify the type of neutrino interaction that produced the event display.

\section{Data analysis}\label{sec:dataprep}

The purpose of this analysis is to use MicroBooNE data and simulation to investigate whether any anomalous neutrino oscillations are present that would be consistent with the existence of a fourth, sterile neutrino state within a $3+1$ flavour model. Such a model could potentially explain the anomalies reported by the LSND and MiniBooNE experiments (see section~\ref{sec:3+1neut}).

\subsection{Analysis strategy}
The neutrino beam used by MicroBooNE is primarily composed of muon neutrinos, as it originates predominantly from charged-pion decays. Over the short baseline between the beam production target and the MicroBooNE detector, any anomalous oscillation involving a sterile neutrino would most likely appear as a {\bf disappearance of muon neutrinos}, quantified by the survival probability $1 - P(\nu_\mu \rightarrow \nu_\mu)$. As a result, CC muon-neutrino interactions, which can be identified by the outgoing muon, are defined as the signal in this analysis and serve as the reference for studying potential disappearance effects.

Initially, students develop a series of event selection criteria to isolate CC muon-neutrino interactions (see section~\ref{sec:cuts}). Once the selected samples are considered to be of sufficiently high purity, the neutrino oscillation hypothesis is tested.

This is first approached by applying the two-flavour muon-neutrino disappearance probability, $1 - P(\nu_\mu \rightarrow \nu_\mu)$, to the Monte Carlo (MC) samples (see section~\ref{sec:TwoFlavourAnalysis}). The oscillation parameters $\theta$ and $\Delta m^2$ are varied, and statistical techniques, as described in section~\ref{sec:appliedtheory}, are used to quantitatively assess how well the predicted distributions describe the data.

Subsequently, students apply the $3+1$ model (section~\ref{sec:ThreeFlavourInterpretation}) to interpret their observations within the broader context of the LSND and MiniBooNE experimental results.

Completion of the steps outlined above constitutes the core component of this laboratory experiment. If students complete this within the allocated time, they are encouraged to extend the analysis in any direction of their choosing. A selection of commonly explored extensions is provided in section~\ref{sec:extend}.

\subsection{Data samples} \label{ssec:topology}
Event displays qualitatively illustrate that different particle species leave characteristic signatures in the detector, enabling a visual classification of interaction topologies. A quantitative analysis, however, requires the  reconstruction of kinematic and topological variables across the entire dataset. Several dedicated software frameworks exist to translate the raw LArTPC information into such higher-level observables; in this work we employ the Pandora reconstruction framework~\cite{MicroBooNE:2017xvs}. Three distinct samples are used in this study to represent different types of interactions:

\begin{itemize}
\item  \textbf{Data:} Reconstructed experimental data collected by the MicroBooNE detector while the neutrino beam was operational.
\item {\bf Monte Carlo (\textbf{MC}) Simulation:} A comprehensive simulation that models the neutrino beam, neutrino–argon interactions, and detector response. The simulation is passed through the same reconstruction algorithms as the data, producing a sample that reproduces the main features of the real detector output.
\item {\bf EXT (External) Sample:} A dataset collected by MicroBooNE when the neutrino beam was off. This is used to characterise backgrounds from non-beam-related activity, such as cosmic rays and electronic noise in the detector.
\end{itemize}

These samples are formatted such that each entry (referred to as an “event”) contains a single neutrino interaction—or background event—occurring within the detector. Each event, in both MC and data samples, contains a collection of reconstructed variables, such as track lengths and total track energy, the latter estimated from the measured  ${\rm d}E/{\rm d}x$.
The MC sample additionally includes ``truth” information from the simulation, providing the true classification of each event according to the interaction categories defined in table~\ref{tab:types}.

\begin{table}[t]
\centering
\begin{tabular}{l|l|l}
\hline
&Interaction & Description \\ \hline
Signal: & {$\nu_\mu $} CC & Charged current muon-neutrino scattering  \\ \hline
Background: & {$\nu_e$} CC & Charged current electron-neutrino 
scattering \\ \hline
& {$\nu $} NC & Neutral current neutrino scattering   \\ \hline
& {Out Fid.\ Vol.} & Neutrino interactions outside the fiducial volume \\ \hline
& {EXT} & Events detected when the neutrino beam is off     \\ \hline
& {Mis-ID} & Cosmic rays misidentified as neutrino events \\ \hline

\end{tabular} 
\caption{Description of the different true event classifications provided in the MC sample}\label{tab:types}
\end{table}

The full list of variables used in this analysis is provided in table~\ref{tab:table_data}. Certain features present in the data, such as cosmic-ray muons and other activity not related to the neutrino beam, are not simulated in the original MC sample. These background processes are modelled using a data-driven technique: data from a subset of the EXT sample---collected when the neutrino beam was off---are overlaid onto the MC sample containing only the beam-initiated neutrino interactions.

The resulting combined dataset, referred to as the \textbf{MC~+~EXT} sample, serves as the background-inclusive prediction. From this point forward, this merged dataset will be referred to simply as the MC sample. This simulated sample does not include any oscillation effects.


\subsection{Selection}\label{sec:cuts}

\begin{figure}[t]
\includegraphics[width=0.49\textwidth]{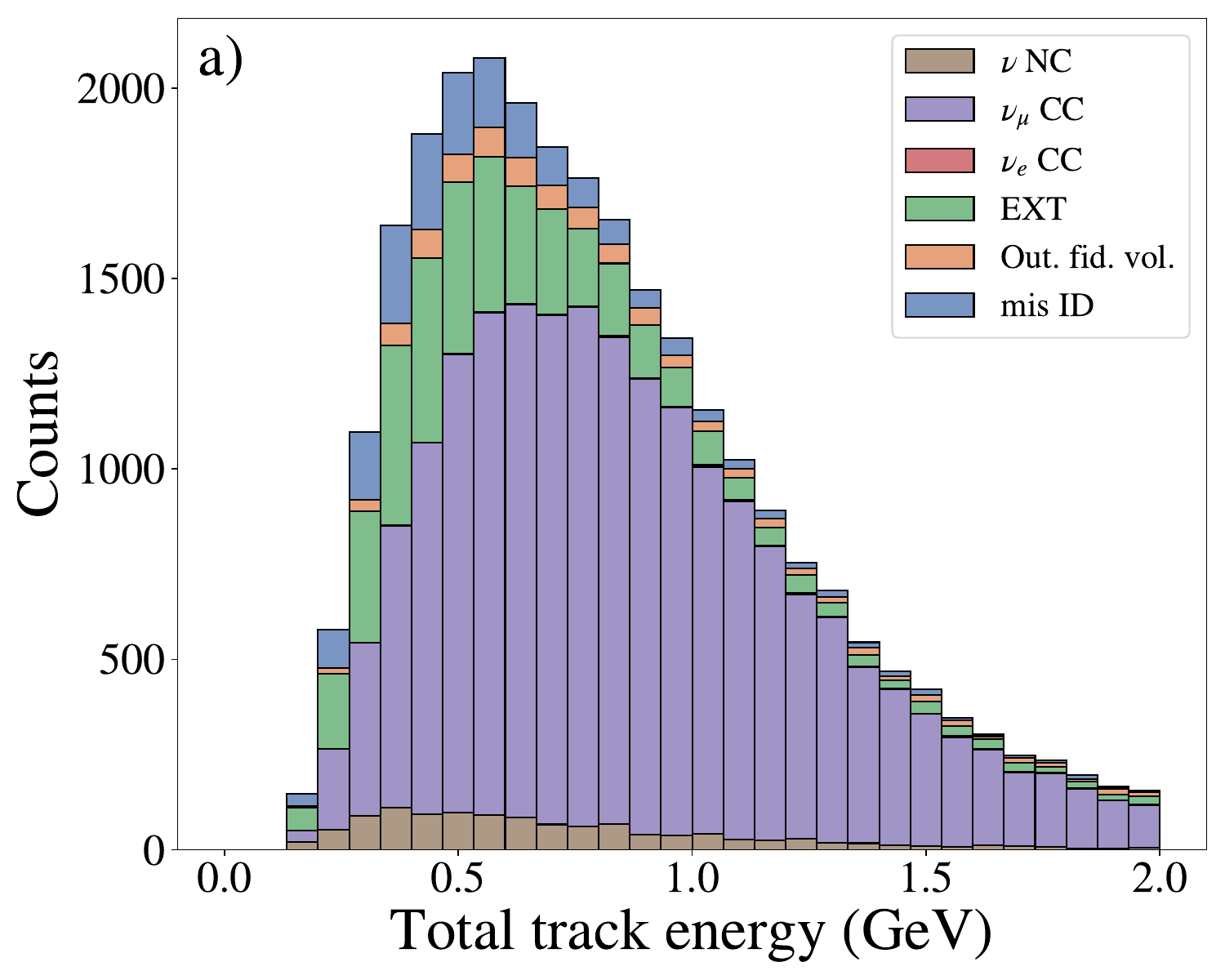}\includegraphics[width=0.49\textwidth]{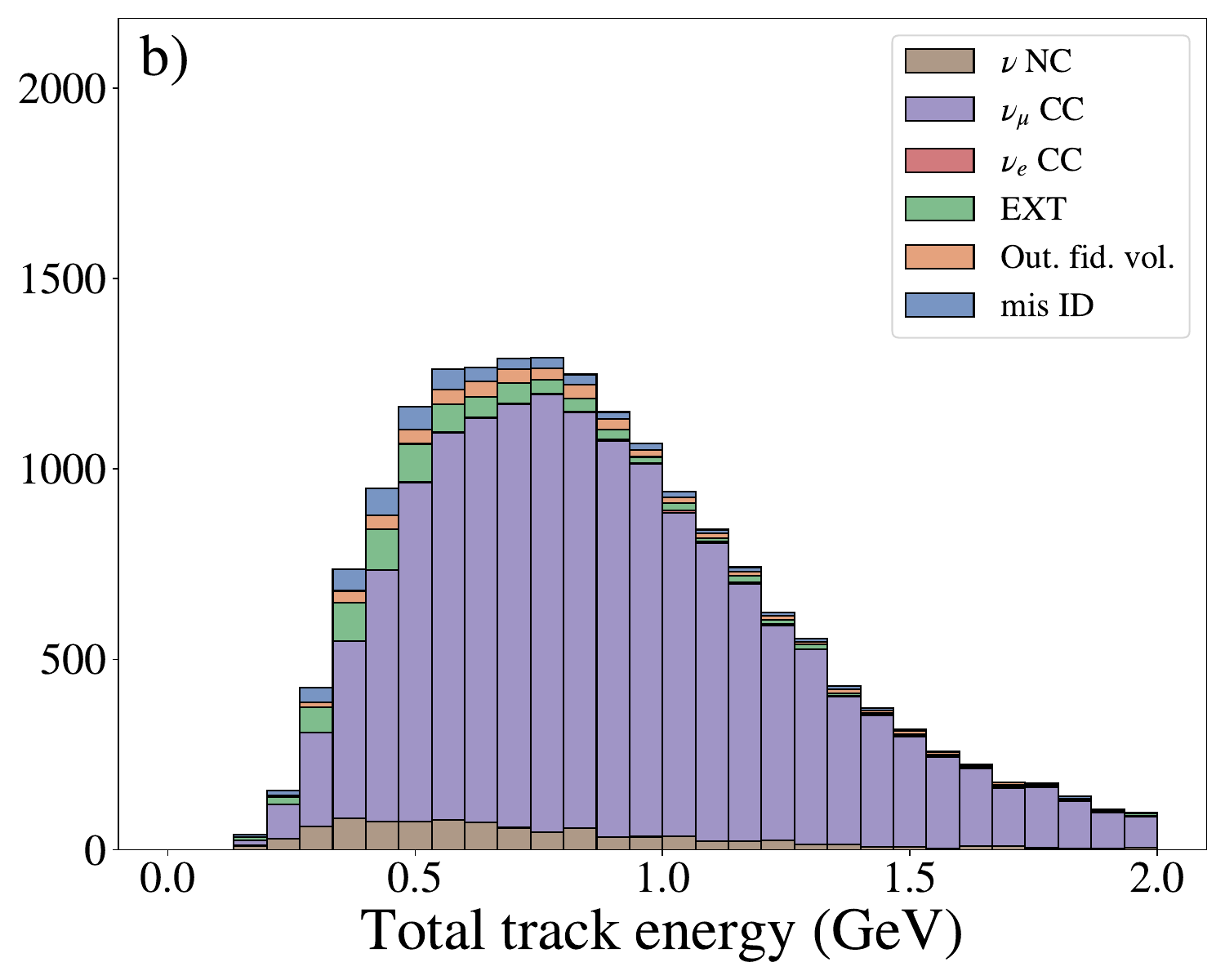}
\includegraphics[width=0.49\textwidth]{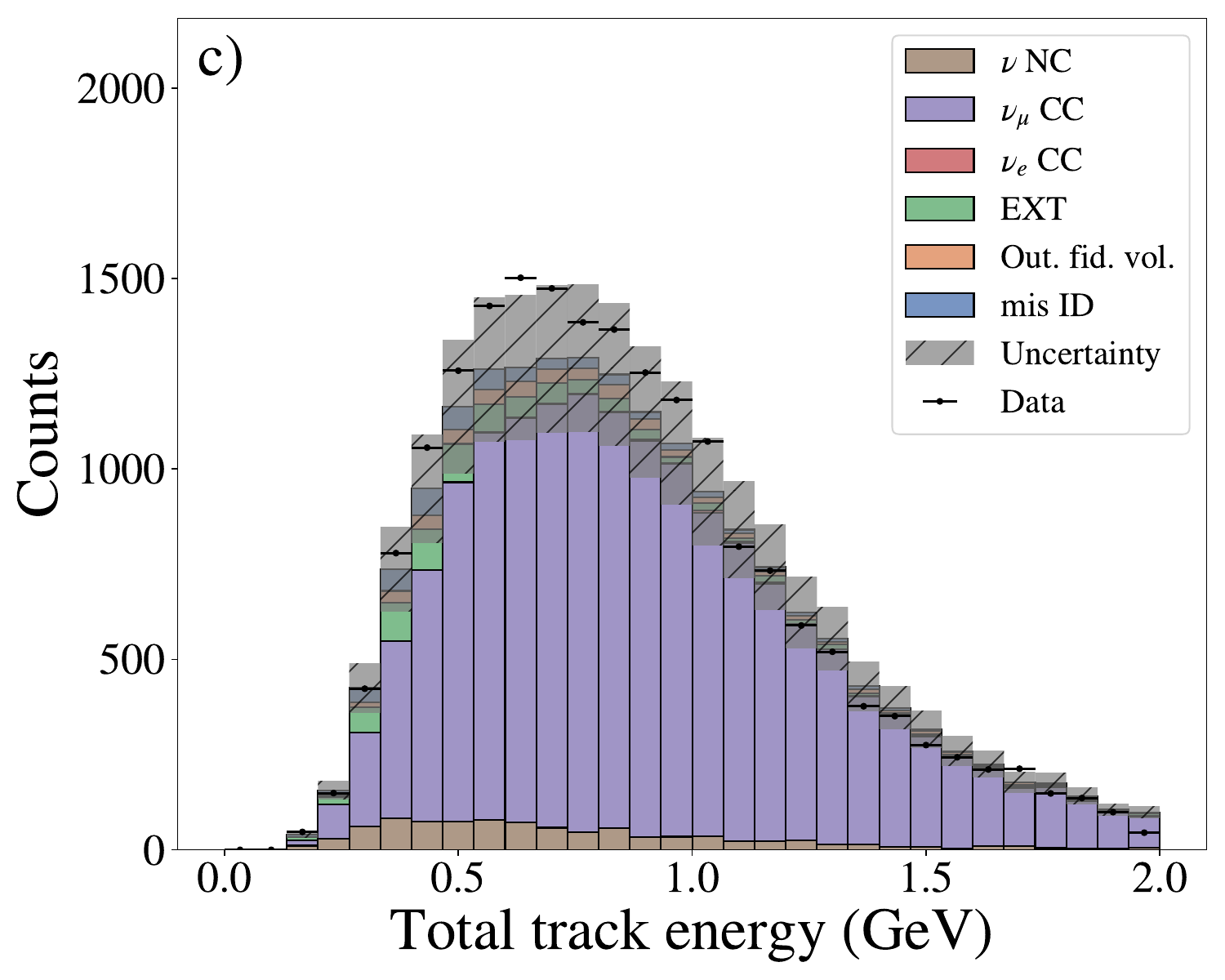}
\includegraphics[width=0.49\textwidth]{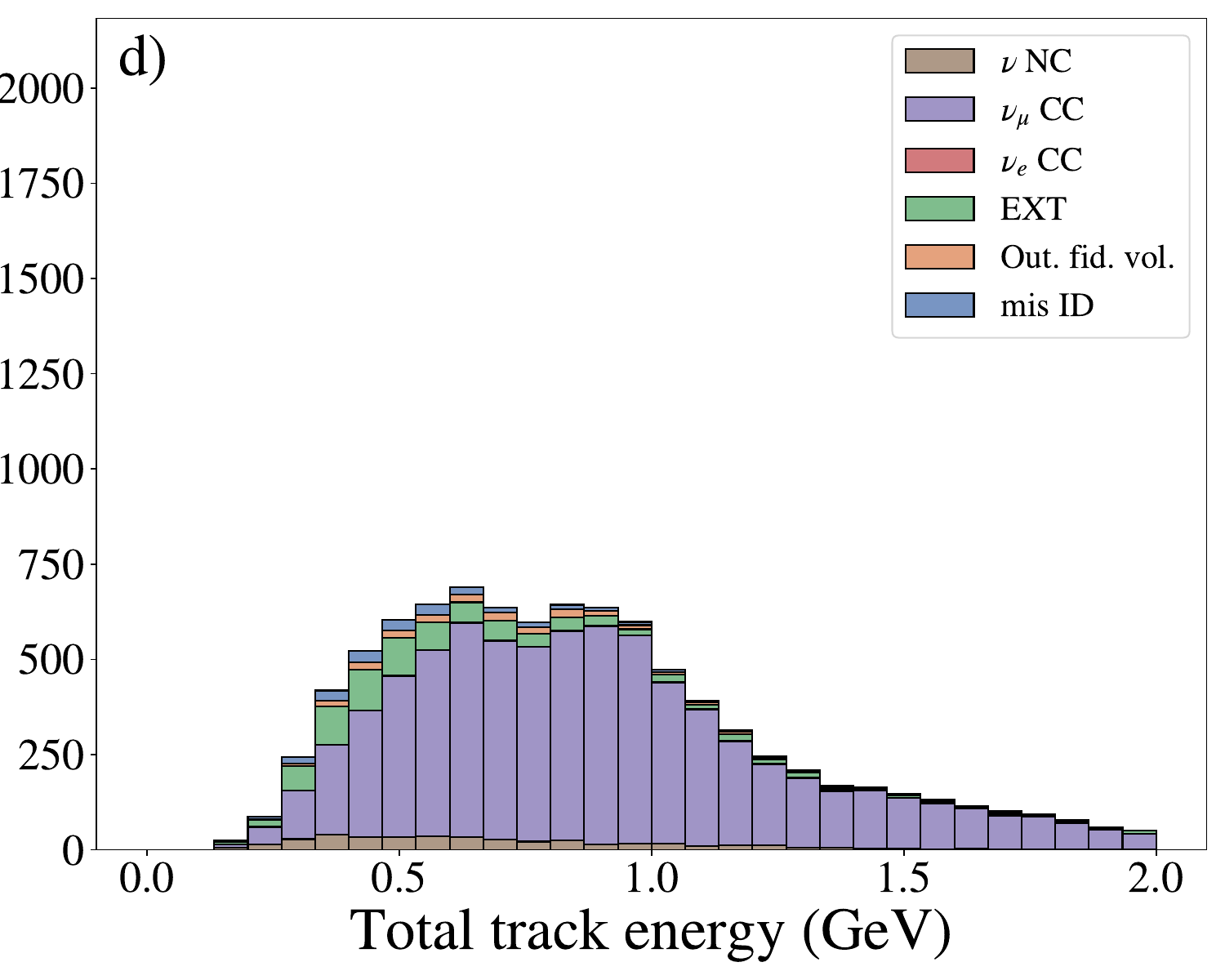}
\caption{The filled histograms show the distributions of events in the MC sample as a function of the total energy summed over all tracks in the event, broken down into the various simulated interaction types:
(a) before any selections have been applied; (b) after a selection has been applied; (c) after selection and
with systematic uncertainties on the MC sample, with the data overlaid;
(d) after the selection, with a two-flavour muon-neutrino disappearance probability applied to the MC sample for $\Delta m^2=\unit[10]{eV^2}$, $\sin^2(2\theta)=0.99$.} 
\label{fig:fullhistos}
\end{figure}

Before considering any oscillation analysis, methods to select CC muon-neutrino events and reduce the number of background events (any other type of event) from the data and MC samples must be applied.
Figure~\ref{fig:fullhistos}(a) shows all the events in the MC sample prior to any background reduction, broken down into their true interaction type.
The $x$ axis of this distribution is the total energy of tracks in each event. The energy of each track is estimated by integrating $\dd E/\dd x$ along its length. The sum of track energies within an event then serves as an estimator for the energy of the incoming neutrino.


In this analysis, a simple sequential selection is applied to topological parameters. The true particle interaction information from the simulation is used to evaluate the performance of the selection. We define two key metrics: \textbf{efficiency}, which quantifies the fraction of signal events retained after the selection, and \textbf{purity}, which measures the fraction of selected events that are true signal events (i.e., the degree of background rejection). These metrics are used to determine the optimal topological selection criteria to be applied to the data.

The metrics are defined as follows in terms of the number of events, $N$, in each category:
\begin{equation}
\text{Efficiency} = \frac{N_{\text{selected signal}}}{N_{\text{total signal}}}, \quad
\text{Purity} = \frac{N_{\text{selected signal}}}{N_{\text{selected total}}}.
\end{equation}
Data analysis tools such as \texttt{seaborn}~\cite{seaborn} are used to determine the final selection, as they allow for visualisation of correlations and distinctions between the different event types described in table \ref{tab:types}.

\begin{table}[t]
  \begin{center}
    
    \begin{tabular}{p{4.8cm}|p{10cm}}
      \hline
      \textbf{Variable Name} & \textbf{Description}\\ 
      \hline
      $\mathrm{vertex_{x,y,z}} \newline \texttt{reco\_nu\_vtx\_sce\_x,y,z}$ & Reconstructed position of the vertex of an interaction in the detector (cm).\\ 
      \hline
      $\mathrm{track_{x,y,z}^{start}} \newline \texttt{trk\_sce\_start\_x,y,z\_v}$ & Reconstructed position of the start of an identified track in the detector (cm).\\ 
      \hline
      $\mathrm{track_{x,y,z}^{end}}  \newline \texttt{trk\_sce\_end\_x,y,z\_v}$ & Reconstructed position of the end of an identified track in the detector (cm).\\ %
      \hline
      $\mathrm{L_{distance}} \newline \texttt{trk\_distance\_v}$ & The distance between the longest track in an event and the reconstructed neutrino vertex (cm).\\ 
      \hline
      $\mathrm{L_{length}} \newline \texttt{trk\_len\_v}$ & The length of the tracks belonging to the detected neutrino interaction (cm).\\ 
      \hline
      $\mathrm{d_{track}} \newline \texttt{trk\_llr\_pid\_score\_v}$ & Log-likelihood score based on calorimetric information used to determine the identity of particles, e.g., muons and protons~\protect\cite{MicroBooNE:2021ddy}.\\
      \hline
      $\mathrm{L_{\nu, cosmic}} \newline \texttt{\_closestNuCosmicDist}$ & Distance between the detected neutrino vertex and closest cosmic-ray muon track within the detector (cm). \\
      \hline
      $\mathrm{S_{topological}} \newline \texttt{topological\_score}$ & Score that determines how much activity in the detector looks like a muon-neutrino interaction. This variable is determined by a machine learning algorithm (Support Vector Machine), trained with neutrino-like events. \\
      \hline
      $\mathrm{S_{track}} \newline \texttt{trk\_score\_v}$ & A score which determines how much activity in the detector looks like a track (instead of a shower). This is also a variable that is determined by machine learning, where the machine has been trained with track-like activity. \\
      \hline
      $\mathrm{E_n} \newline \texttt{trk\_energy\_tot}$ & 
      Sum of the reconstructed energies of tracks in the event, which is an estimator of the energy of the incoming neutrino (GeV). \\
      \hline
      $\mathrm{P_{\mu}^{range}} \newline \texttt{trk\_range\_muon\_mom\_v}$ & Muon momentum determined by the range (or length) of the muon track through the detector (GeV). \\
      \hline
      $\mathrm{P_{\mu}^{mcs}} \newline \texttt{trk\_mcs\_muon\_mom\_v}$ & Momentum determined from reconstructing multiple Coulomb scatterings of the muon (GeV). \\
      \hline
    \end{tabular}  \caption{A list of variables, and their names within the data structure, available within the MicroBooNE simulation and data samples.}\label{tab:table_data}
  \end{center}
\end{table}

\begin{figure}[t]
    \centering
    \includegraphics[width=0.6\linewidth]{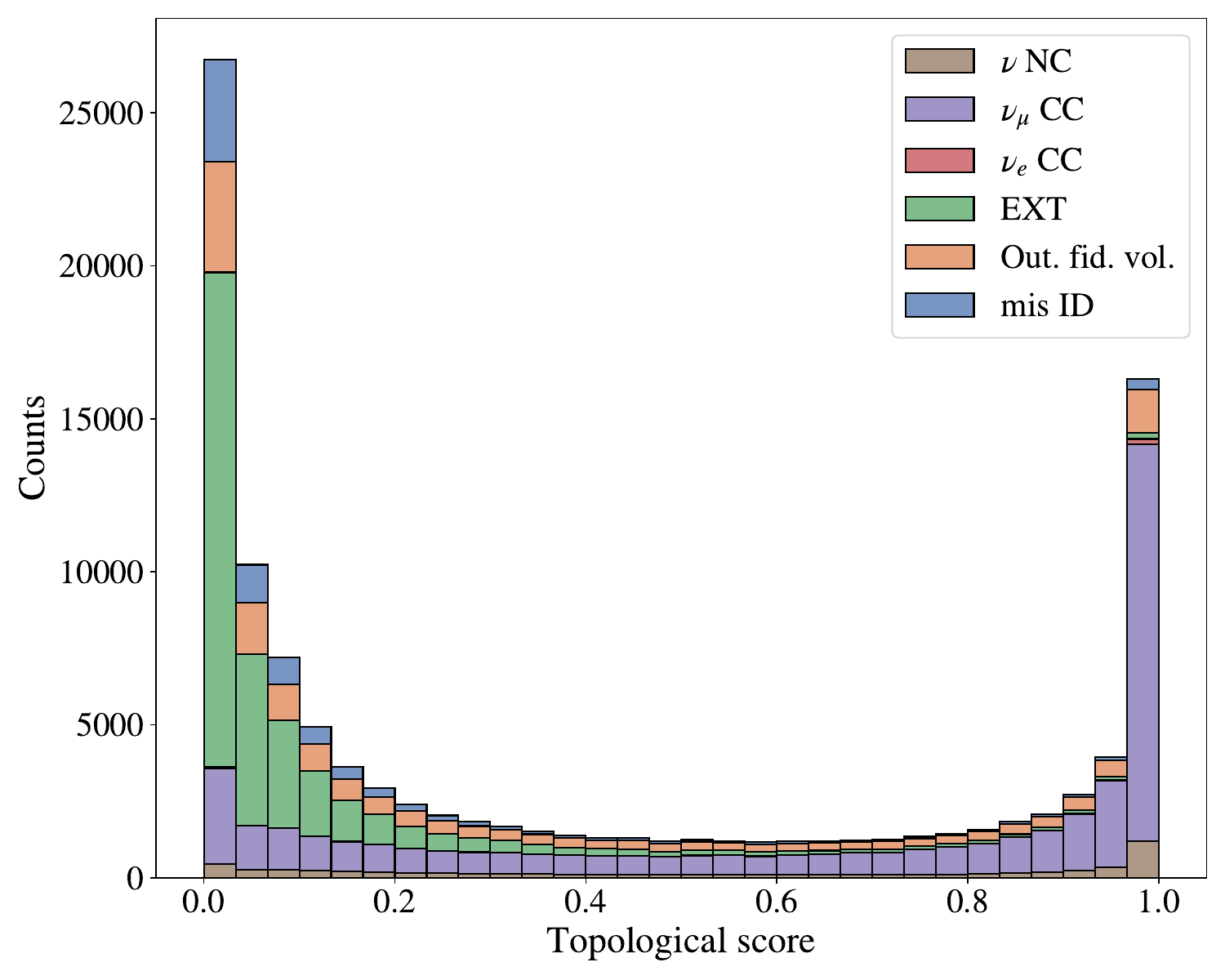}
    \caption{The variable \texttt{topological\_score}, which determines the extent to which activity in the detector looks like a muon-neutrino interaction.}
    \label{fig:topological_score}
\end{figure}

The topological variables available to students are listed in table~\ref{tab:table_data}. The ability of individual variables to discriminate between signal and background events can be evaluated by plotting their distributions for the simulation, separated by interaction type.
An example is shown in figure~\ref{fig:topological_score}, 
which displays the distribution of the variable \texttt{topological\_score}. 
This variable is derived from a machine learning algorithm and quantifies how closely an event resembles a CC muon-neutrino interaction. Applying a threshold on this variable, selecting events with a value greater than, e.g., $0.4$, effectively suppresses background.


\begin{table}[t]
  \begin{center}
    \begin{tabular}{p{5cm}|p{10cm}}
      \hline
      \textbf{Variable Name} & \textbf{Description}\\ 
      \hline
      $\mathrm{category} \newline \texttt{category}$ & Interaction type of the event.\\ 
      \hline
      $\mathrm{weight} \newline \texttt{weight}$ & Weighting of each event.\\ 
      \hline
      $\mathrm{L_{true}}  \newline \texttt{true\_L}$ & True distance between beam interaction point and interaction vertex in detector (cm).\\ %
      \hline
      $\mathrm{E_{true}} \newline \texttt{true\_E}$ & True energy of the neutrino (GeV).\\ 
      \hline
      $\mathrm{P^{true}_\mu} \newline \texttt{true\_muon\_mon}$ & True muon momentum (GeV).\\ 
      \hline
\end{tabular}  \caption{A list of all MC variables, and their names within the data structure, available within the MicroBooNE simulation sample.}\label{tab:MC_table_data}
  \end{center}

\end{table}

Students are encouraged to assess the discriminating power of the available variables, select those most suitable for event selection, and use data visualisation tools to investigate correlations between variables to further optimise the selection requirements. Through this process, students can typically achieve an efficiency of approximately $80\%$ and a purity of around $85\%$.

Figure~\ref{fig:fullhistos}b shows the distribution of the total energy of tracks in selected events following the application of a typical event selection. In comparison to figure~\ref{fig:fullhistos}a, prior to the application of the selection criteria, an improvement in sample purity is observed, providing a more suitable dataset for further analysis. For the remainder of the laboratory, students continue their work using this selected event sample.

\subsection{Decision trees} \label{sec:dectrees}

When analysing large datasets involving numerous variables, as in this study, identifying patterns or correlations becomes increasingly challenging. This is where the application of ML techniques can enhance event classification. 
ML methods provide tools to explore more complex relationships among variables and to leverage this information in order to achieve improved separation between signal and background events.

\textbf{Decision trees} are employed within the MicroBooNE collaboration for various tasks, most notably for the rejection of background events~\cite{decision_t1, decision_t2, decision_t3}.
After developing this sequential selection based on individual variables, students are introduced to the use of ML as an alternative tool for event classification in large datasets by implementing decision trees.
The \textbf{Gini criterion} is employed as the splitting metric to evaluate the effectiveness of the classifier, and \textbf{confusion matrices} are used to assess the overall performance of their model.

Decision trees are prone to \textbf{overfitting}, whereby a model becomes overly tailored to the training dataset, resulting in a loss of generalisability and reduced accuracy when applied to independent samples from the same simulation. To mitigate this issue, the use of \textbf{random decision forests} implemented in the \texttt{scikit-learn} package~\cite{scikit} is recommended for student use.

Once students have implemented their ML model, they should compare the purity and efficiency with that achieved by their sequential selection and discuss the comparison.
Typically their ML model does not achieve a significant improvement, since some of the variables used in the sequential selection, as described in table~\ref{tab:table_data}, are already determined by ML.


\section{Hypothesis and model testing}\label{sec:appliedtheory}
In this section, we discuss several statistical tools applied to facilitate the quantitative testing of the $3+1$ sterile neutrino model using the available data.

\subsection{Statistical and systematic uncertainties} \label{sec:unc}

Identical selection criteria are applied to the simulated MC dataset to enable direct comparison with experimental data. In order to allow for meaningful comparisons between the data and MC distributions, both statistical and systematic uncertainties must be quantified.

The statistical uncertainty associated with the data histogram arises from inherent randomness in sampling the underlying distribution. The uncertainty in each bin is taken to be $\sqrt{N}$, where $N$ denotes the expected number of events in that bin, derived from the simulation, in accordance with the standard deviation of a normal or Gaussian distribution~\cite{cowan_statistical_1998}.

In contrast, estimating systematic uncertainties is substantially more complex. A realistic evaluation must account, at a minimum, for uncertainties in neutrino-beam flux modelling, neutrino--argon interaction cross-sections, and detector response.
As a full treatment of these uncertainties~\cite{many_universe_uboone, multisim} lies beyond the scope of this laboratory, a fixed relative uncertainty of $15\%$ is applied as an approximation, assumed to be uncorrelated between bins. 
This is combined in quadrature with the statistical uncertainty, as illustrated in figure~\ref{fig:fullhistos}(c).

\subsection{Confidence levels} \label{sec:oscil]}

We want to quantify whether the experimental data are consistent with the null hypothesis (no oscillations at this baseline, $L$) or an alternative hypothesis (in this case a two-flavour or $3+1$ oscillation model with a sterile neutrino).
The task of deciding between different models is common in statistics and is known as a
hypothesis test.  The goal
of the hypothesis test is to quantify the agreement of the experimental data
with the prediction of each hypothesis.

We perform this hypothesis test by comparing
the simulated distributions of the total reconstructed track energy, obtained under different hypotheses of values for the mass splitting and mixing angle, to the equivalent distribution from the data, as illustrated in figure~\ref{fig:fullhistos}c.
The goodness of the fit of the hypothesis is quantified using a $\chi^2$ metric,
\begin{equation}
\chi^2 = \sum_{i=1} \left(\frac{(\mu_{i} - M_{i})^{2}}{\sigma_{i}^2}\right), 
\end{equation}
 where $\mu_i$ is the number of events in the $i$th bin of the oscillated MC histogram, $M_i$ is the number of events in the $i$th bin in the data histogram, and $\sigma_i$ is the uncertainty across each bin as described in section~\ref{sec:unc}. 

 We express the consistency of the data with a given hypothesis using a \textbf{confidence level} (CL), which is derived from the $\chi^2$ distribution and the associated $p$-values. For example, in a simple Gaussian case with one parameter, a range of $\pm 1$ standard deviation corresponds to a CL of $68\%$ and a $p$-value of $0.32$.

These probabilities depend on the model parameters (or nuisance parameters)---in our case, $\Delta m^{2}$ and $\sin^{2}(2\theta)$. As the model under consideration has two parameters, the two-dimensional CL contours are constructed by identifying the boundary in parameter space at which the difference in $\chi^2$, denoted $\Delta\chi^2$, between a point and the minimum $\chi^2$ value reaches a critical threshold.
The \textbf{critical values} of $\Delta\chi^2$ corresponding to specific confidence levels for two-parameter models are obtained from standard statistical reference tables~\cite{pdg}.

In this \textbf{frequentist} framework, the interpretation is based on the outcome of a large ensemble of hypothetical (``gedanken'') experiments. The CL represents the probability of observing data at at least as extreme a value as the actual dataset, assuming a particular hypothesis is true. By convention, the results of hypothesis tests are typically reported at the \unit[90\%]{CL} or \unit[95\%]{CL}.

 \subsection{Reweighting}
The MC simulation does not incorporate neutrino oscillations by default. However, different oscillation hypotheses can be evaluated by comparing oscillated MC predictions with the experimental data. To simulate the effects of a potential neutrino oscillation, an oscillation probability is computed for each neutrino event in the MC sample.

This probability depends on the mass-squared splitting, the mixing angle, the true neutrino energy $E$, and the distance travelled, $L$ (see equations~(\ref{eq:natty}) and~(\ref{eq:TwoFlavourSurvivalProb})). The probability is applied to each simulated event as a weight, thereby modifying the predicted distribution to reflect the specified oscillation parameters.

An example of such an oscillated prediction---where a muon-neutrino disappearance probability has been applied---is shown in figure~\ref{fig:fullhistos}d.

\section{Oscillation analysis}

\subsection{Two-flavour muon-neutrino disappearance}\label{sec:TwoFlavourAnalysis}

Students employ the two-flavour model to investigate potential evidence for muon-neutrino disappearance. This is achieved by applying the survival probability $P(\nu_\mu \rightarrow \nu_\mu)$ from equation~(\ref{eq:TwoFlavourSurvivalProb}) to weight the MC prediction, thereby simulating the effect of oscillation-induced disappearance (see figure~\ref{fig:fullhistos}d).
Equation~(\ref{eq:TwoFlavourSurvivalProb}) contains two free parameters: $\Delta m^2$ and $\sin^2(2\theta)$. Varying these parameters alters the MC prediction and, consequently, the $\chi^{2}$ value when the simulation is compared to the experimental data.

\begin{figure}[ht] 
\centering
\includegraphics[width=0.6\textwidth]{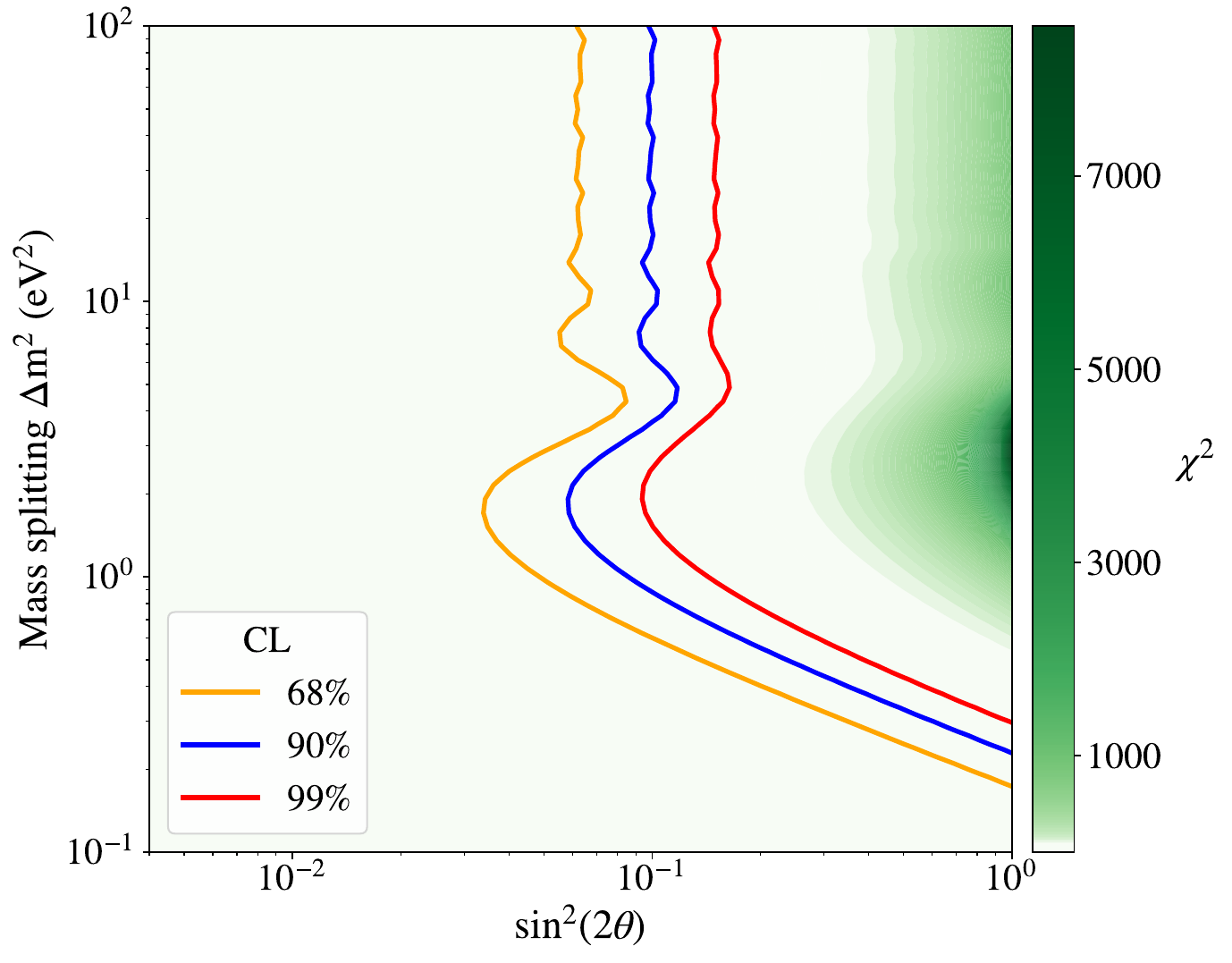}
\caption{Visualisation of $\chi^2$ values mapped across the parameter space of the two-flavour oscillation model from a muon-neutrino disappearance search.
This mapping highlights the \((\Delta m^{2},\sin^{2} (2\theta))\) regions where the oscillated MC prediction is statistically incompatible with the data set.
The confidence-level (CL) contours delineate the regions of parameter space disfavoured by the data.
Any point lying to the right of a given contour is excluded at that confidence level.} \label{fig:chisquare}
\end{figure}

By systematically varying the parameters across a grid of values, students construct a two-dimensional map of $\chi^2$ values over the $(\Delta m^2,\, \sin^2(2\theta))$ parameter space, as shown in figure~\ref{fig:chisquare}. The best-fit values of $\Delta m^2$ and $\sin^2(2\theta)$ correspond to the point at which the $\chi^2$ reaches its minimum.

When comparing the non-oscillated MC and data distributions, the MC prediction lies below the corresponding data histogram in nearly every bin, as shown in figure~\ref{fig:fullhistos}(c). Since muon-neutrino disappearance uniformly reduces the event weights in the MC sample, any oscillated MC distribution will also remain below the data.
As a result, obtaining an ideal $\chi^2$ value that indicates a good fit is not feasible in this case. Consequently, the best-fit oscillation parameters will typically be $(\Delta m^{2}, \sin^{2} (2\theta)) = (0, 0)$.

Rather than seeking an optimal fit, the analysis focuses on evaluating how unlikely it is that an oscillated MC sample---generated with specific values of $\Delta m^2$ and $\sin^2(2\theta)$---can accurately describe the MicroBooNE data. Larger values of $\chi^2$ indicate poorer agreement with the observed distribution and, therefore, enable certain regions of the oscillation parameter space to be \textbf{excluded} as inconsistent with the measured neutrino energy spectrum.

This inconsistency is quantitatively expressed using CLs, as shown in figure~\ref{fig:chisquare} which displays regions of parameter space excluded at \unit[68\%]{CL}, \unit[90\%]{CL}, and \unit[95\%]{CL}. These contours are derived using the critical values from the statistical reference tables in ref.~\cite{pdg}.

\subsection{Expressing results in a (3+1)-flavour model} \label{sec:ThreeFlavourInterpretation}

The two-flavour analysis carried out by the students quantifies whether there is any evidence for muon-neutrino disappearance. However, it does not allow for a direct comparison with the MiniBooNE and LSND observations, as the two-flavour model does not specify into which non-muon flavour states the neutrinos are oscillating.
In the notation of section~\ref{sec:TwoFlavourModel}, the two-flavour model describes the transition $\nu_\mu \rightarrow \nu_\beta$, where $\nu_\beta$ may correspond to any of the three other flavour states---or indeed to an admixture of them.

To enable students to interpret their results in the context of the LSND and MiniBooNE observations, a $3+1$ model must be employed. In this extended framework, the flavour-change probability described in equation~(\ref{eq:natty}) can be decomposed into components that quantify the extent to which muon neutrinos oscillate into each of the other three flavour states: $\nu_e$, $\nu_\tau$, and $\nu_s$. The corresponding flavour-change probabilities are given by:
\begin{eqnarray}
    \hspace{-1.5cm}P(\nu_\mu \rightarrow \nu_\beta) &\hspace{0cm}=& \hspace{0cm}\sin^2(2\theta_{\mu\beta})\sin^2\left(1.27\frac{\Delta m^2_{41}(\textrm{eV}^2) L(\textrm{km})}{E(\textrm{GeV})}\right),\qquad\beta=(e,\tau,s),\\
    \hspace{-1.5cm}P(\nu_\mu\rightarrow\nu_\mu)&=&1-\sin^2(2\theta_{\mu\mu})\sin^2\left(1.27\frac{\Delta m^2_{41}(\textrm{eV}^2) L(\textrm{km})}{E(\textrm{GeV})}\right).\label{eq:ThreeFlavourSurvivalProb}
\end{eqnarray}
We define four new effective mixing-angle parameters: $\theta_{\mu e}$, $\theta_{\mu\mu}$, $\theta_{\mu\tau}$, and $\theta_{\mu s}$. The relationship between these and the underlying parameters $\theta_{ij}$ of the $4 \times 4$ PMNS matrix, as introduced in equation~(\ref{eqn: PMNS_4_flavour}), is detailed in the literature (see, for example, refs.~\cite{ref:GiuntiParameterisation, ref:HarariParameterisation}).

For the purposes of this experiment, we are concerned only with the $\nu_\mu \rightarrow \nu_\mu$ and $\nu_\mu \rightarrow \nu_e$ transitions. As a result, the only relevant parameters are:
\begin{eqnarray}
    \sin^2(2\theta_{\mu e})&=&\sin^2(2\theta_{14})\sin^2\theta_{24},\label{eq:ThetaMuEDefinition}\\
    \sin^2(2\theta_{\mu\mu})&=&4\cos^2\theta_{14}\sin^2\theta_{24}(1-\cos^2\theta_{14}\sin^2\theta_{24}).\label{eq:ThetaMuMuDefinition}
\end{eqnarray}
By comparing equations~(\ref{eq:TwoFlavourSurvivalProb}) and~(\ref{eq:ThreeFlavourSurvivalProb}), it can be seen that the parameter $\sin^2(2\theta)$ used in the two-flavour muon-neutrino disappearance analysis corresponds directly to $\sin^2(2\theta_{\mu\mu})$ in the $3+1$ model.
Therefore, if a value of $\theta_{14}$ is adopted from an external source, students may use equations~(\ref{eq:ThetaMuEDefinition}) and~(\ref{eq:ThetaMuMuDefinition}) to convert their constraints on $\theta_{\mu\mu}$ into corresponding constraints on $\theta_{\mu e}$. These can then be directly compared with the LSND and MiniBooNE observations.

Students are asked to use the expressions above to derive the following relationship:
\begin{equation}
\sin^2 (2\theta_{\mu e}) = \left(1-\sqrt{1-\sin^2 (2\theta_{\mu \mu})}\right)\left(1-\sqrt{1-\sin^2 (2\theta_{ee})}\right).
\label{eq:sinue}
\end{equation}
As a representative value, we adopt $\sin^2(2\theta_{14}) = 0.24$, which corresponds to the best-fit point from a previous analysis~\cite{ref:MicroBooNESterile}. It should be noted, however, that in that analysis $\theta_{14}$ was found to be consistent with zero at the \unit[95\%]{CL}.

\begin{figure}[ht] 
\centering
\includegraphics[width=0.7\textwidth]{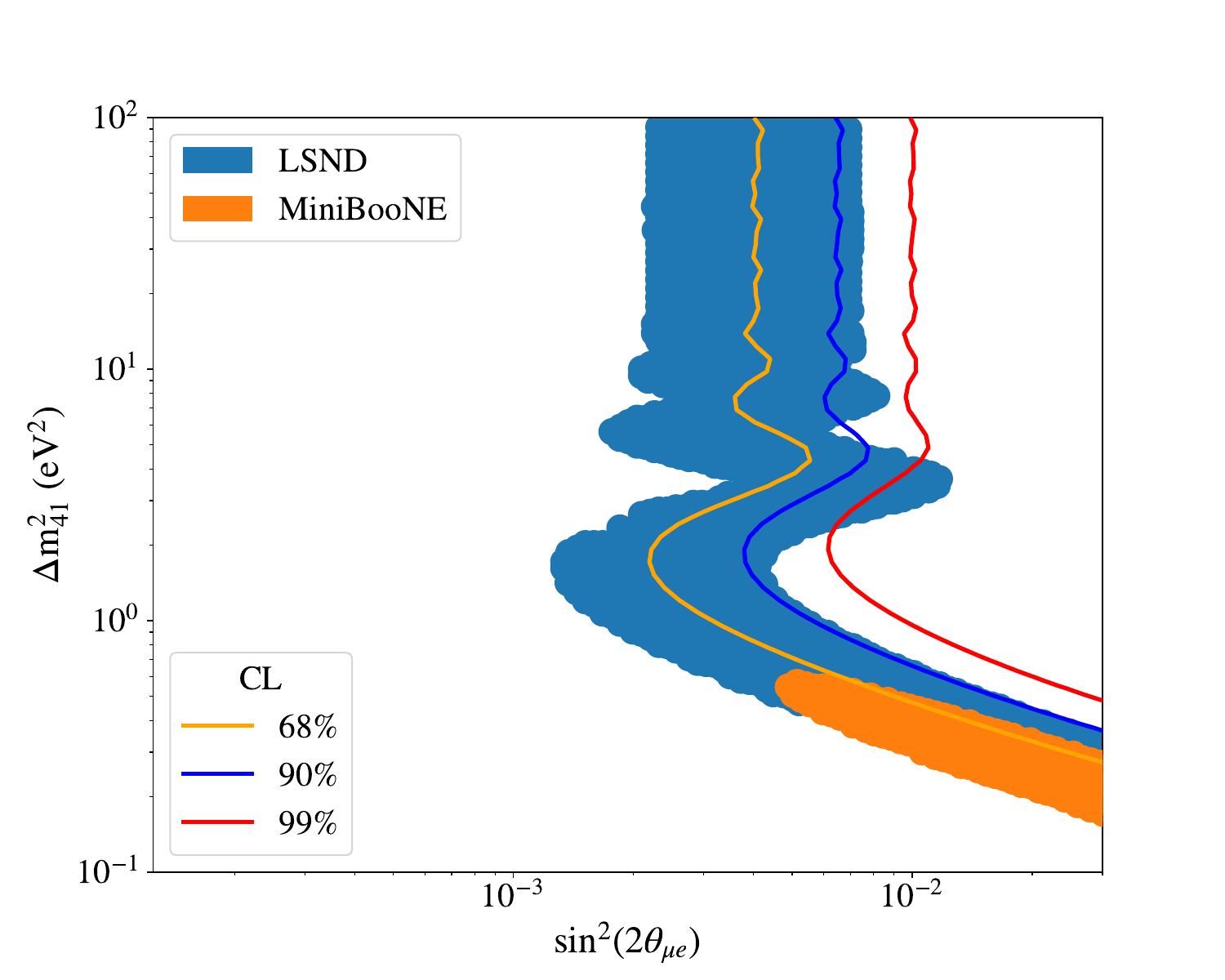}
\caption{The lines are the exclusion contours obtained from the $\chi^2$ map of figure~\ref{fig:chisquare}, converted into the parameter space of the $3+1$ model using a value of $\sin^2(2\theta_{14})=0.24$.
The regions to the right of these lines are excluded at the given confidence level (CL).
These exclusion contours are compared to the regions of the parameter space favoured, at \unit[90\%]{CL}, by the LSND and MiniBooNE observations in the $3+1$ model~\cite{lsndminibooneCLs}.}
\label{fig:chi_squared_lsnd_miniboone}
\end{figure}

Figure~\ref{fig:chi_squared_lsnd_miniboone} shows the result when the $\chi^2$ map of figure~\ref{fig:chisquare} is converted into the $(\Delta m^2_{41},\sin^2(2\theta_{\mu e}))$ parameter space of the $3+1$ model using a value of $\sin^2(2\theta_{14})=0.24$.
For clarity, the converted $\chi^2$ surface is not itself shown, but the exclusion contours obtained using the \unit[68\%]{CL}, \unit[90\%]{CL} and \unit[99\%]{CL} critical values with respect to the minimum $\chi^2$ value are shown.
These are compared to the \unit[90\%]{CL} allowed regions from the MiniBooNE and LSND analyses.
 
Figure~\ref{fig:chi_squared_lsnd_miniboone} presents the result of converting the $\chi^2$ map from figure~\ref{fig:chisquare} into the $(\Delta m^2_{41}, \sin^2(2\theta_{\mu e}))$ parameter space of the $3+1$ model, using the representative value $\sin^2(2\theta_{14}) = 0.24$.
For clarity, the full converted $\chi^2$ surface is not shown; instead, the exclusion contours corresponding to the \unit[68\%]{CL}, \unit[90\%]{CL}, and \unit[99\%]{CL} are plotted, based on critical values relative to the minimum $\chi^2$ value. These exclusion regions are compared to the \unit[90\%]{CL} allowed regions from the MiniBooNE and LSND analyses.

\subsection{Interpreting the results}\label{sec:results}

The exclusion contours shown in figure~\ref{fig:chi_squared_lsnd_miniboone} within the $3+1$ model allow a direct comparison between the students’ results and the observations reported by MiniBooNE and LSND. The corresponding regions in figure~\ref{fig:chi_squared_lsnd_miniboone} represent the \unit[90\%]{CL} areas within which the existence of a sterile neutrino with the corresponding $\Delta m^2_{41}$ and $\sin^2(2\theta_{\mu e})$ values could account for the anomalies observed by LSND and MiniBooNE~\cite{lsndminibooneCLs}.

Only a portion of the LSND-allowed region---and none of the MiniBooNE-allowed region---is excluded at the \unit[90\%]{CL} by this simplified ``toy" MicroBooNE analysis. This outcome demonstrates that the toy analysis is not sufficiently sensitive to probe the majority of the LSND and MiniBooNE parameter space. Students should therefore conclude that the data disfavours the $3+1$ sterile-neutrino model; however, it remains a viable explanation given the limited sensitivity of the analysis.

The MicroBooNE collaboration investigates the short-baseline neutrino anomaly primarily through searches for \textbf{electron-neutrino appearance}~\cite{ref:MicroBooNESterile}, mirroring the approach adopted by LSND and MiniBooNE. In contrast, the analysis described in this paper focuses on a \textbf{muon-neutrino disappearance} search.

From a pedagogical standpoint, this difference in strategy offers a valuable opportunity to discuss parameter inference across distinct experimental methodologies. It invites students to consider how identical physical parameters---such as $\sin^2(2\theta_{\mu e})$ and $\Delta m^2_{41}$---can be extracted using different neutrino flavours and experimental techniques.

\section{Extensions}\label{sec:extend}
If students complete all the previous work outlined in this paper within the allotted time, they are encouraged to improve upon their work. These improvements can be categorised into three general areas: enhancing their results (\textbf{optimise}), developing a better understanding of the results (\textbf{understand}), and testing novel or alternative theories and ideas based on this understanding (\textbf{hypothesise}). 

\subsection{Optimisation}

One method for optimising results is by improving the selection criteria applied.
There are various approaches to this optimisation, ranging from simple parameter scans to more complex genetic algorithms---both of which have been employed by students in recent years.
This is an area where ML techniques can be particularly effective in enhancing event classification.
Students are encouraged to explore this avenue both to improve their results and to deepen their understanding of machine learning.

However, given the samples provided, there is a limit to how much optimisation can realistically be achieved.
This becomes most evident when the selection criteria reduce the statistical sample to the point where statistical uncertainties dominate, thereby weakening the exclusion power of the results.

A common issue in all histogram-based data is the prevalence of over- or under-binning. Since the $\chi^2$ values derived from oscillation theory depend directly on the chosen binning scheme, an elegant approach to improving sensitivity is to optimise the number of bins across the data samples using Shannon entropy~\cite{watts2022shannonentropyhistogram}.

\subsection{Understanding}

A full treatment of MicroBooNE’s systematic uncertainties lies beyond the scope of this exercise. Therefore, the histograms used in the statistical analysis are assigned a uniform $15\%$ uncertainty---an ideal starting point for a more in-depth extension focused on systematic studies.
The primary sources of uncertainty in MicroBooNE fall into three broad categories: the neutrino flux simulation, performed using the \textit{GEANT4} package~\cite{GEANT4:2002zbu}; the modelling of neutrino–nucleus interactions by the \textit{GENIE} simulation code~\cite{Andreopoulos:2009rq}; and systematic effects associated with the detector response.
Investigating how and why these uncertainties are determined offers valuable insight into the operation of experiments like MicroBooNE and enables students to engage with a non-trivial area of research in physics.

\subsection{New hypotheses}

Understanding how the uncertainties are determined is useful for familiarising oneself with the inner workings of the experiment, but how would improving them affect the results? 
Hypothesising about differing conditions of the apparatus is a good way to improve the understanding of the underlying physics within this analysis.
If a (hypothetical) near detector could constrain the neutrino flux and in turn reduce the overall uncertainty to 5\%, how would that affect the resultant parameter scan?
This sort of thinking can be applied to all avenues under which systematic uncertainties are considered.

A further extension is to explore more complex oscillation models—--e.g.\ the $3+2$ sterile-neutrino model~\cite{krolikowski2005simplest32modellight}. Testing such alternative hypotheses against the data deepens one’s grasp of neutrino oscillation theory and consequently the field as a whole.

\section{Conclusion and outlook}\label{sec:conclusion}

We have presented an experiment designed for an advanced undergraduate teaching laboratory that enables students to experience research in particle physics, using data from the MicroBooNE liquid-argon time projection chamber.
By the end of the laboratory, students will have developed a method of extracting signal from a large dataset, have had experience with ML, and will have used statistical techniques to compare their data to a MC simulation, incorporating statistical and systematic uncertainties, to set limits on the parameters of a physical model.
Students will have compared their results with equivalent results from the field to draw inferences on the existence of a possible fourth neutrino state.
This experiment also provides students with the opportunity to independently optimise and extend their analysis methods.

Details of how to access the data samples, simulation samples, and software for this laboratory were provided in section~\ref{sec:Tech}.
The implementation of the laboratory described here is designed for students to work for twelve days over six weeks, but it can be easily adapted to different laboratory structures by using only subsets of the activities described, or by incorporating extensions.



\appendix

\section*{Acknowledgements}

We gratefully acknowledge the MicroBooNE Collaboration for making their data available for this project. The samples used in this study are simplified datasets derived from MicroBooNE, intended solely for teaching and pedagogical purposes. They are not representative of full detector performance and are not intended to support physics conclusions. We thank the MicroBooNE Collaboration for providing this resource.
This material is based upon work supported by STFC grant ST/W003945/1.

\newpage
\section*{References}
\bibliography{references}
\end{document}